\documentclass[aps,pra,preprint,amsmath,amssymb,groupedaddress]{revtex4-1}

\usepackage{appendix}
\usepackage{hyperref}
\usepackage{graphicx}

\numberwithin{equation}{section}
\usepackage[english]{babel}

\DeclareMathOperator{\Tr}{Tr}
\usepackage{ mathrsfs }
\usepackage{bbold}
\usepackage{dsfont}
\usepackage{ braket }

\usepackage{physics}
\usepackage{wrapfig}
\usepackage{pifont}
\newenvironment{psmallmatrix}
  {\left(\begin{smallmatrix}}
  {\end{smallmatrix}\right)}

\begin{document}


\title{Discrimination of Quantum States under Local Operations and Classical Commuication}


\author{\"{O}zen\c{c} G\"{u}ng\"{o}r}
\email[]{oxg34@case.edu}
\altaffiliation{Department of Physics, Middle East Technical University, 06800, Ankara, Turkey}
\affiliation{Department of Physics, Case Western Reserve University,
10900 Euclid Avenue, Cleveland, Ohio 44106, USA}


\date{\today}

\begin{abstract}
The problem of quantum state discrimination, which is a foundational aspect of quantum information theory, and its relation to the theory of majorization are discussed. The purpose of this study is to review different approaches to the problem and analyze different cases of quantum-state discrimination, most importantly the discrimination of bipartite entangled quantum states under local operations and classical communication. Two partial results on using entanglement as a resource for quantum-state discrimination and discrimination with remaining entanglement is given. The important points are also summarized and the results are discussed.
\end{abstract}


\maketitle

\tableofcontents

\section{Introduction}
One of the biggest paradigm shifts in the history of science is undoubtedly the Quantum revolution and along with it came many unanswered questions and \textquotedblleft paradoxes\textquotedblright$\,$ and scientists and philosophers alike debated the very nature of reality as quantum mechanics challenged the old, classical ideas. The most famous of these debates are the Bohr-Einstein debates and the papers of EPR (Einstein-Podolsky-Rosen) \cite{EPR1}, Bohr \cite{EPR2}, Bohm and Aharonov \cite{EPR3}. It was in the first EPR paper that the notion that is nowadays known as quantum entanglement was introduced. These papers explored the fundamental concept of physical reality and in a manner, paved the way for J. S. Bell's famous papers \cite{BELL} on the nature of physical reality and the pivotal role quantum entanglement plays. It was after the works of J. S. Bell and his collaborators that researchers began to see entanglement as something other than just a peculiarity of quantum mechanics and started to question the nature of information and its physical meaning. The theory of quantum information was born.  

Quantum information theory explores the physical nature of information and investigates information processing tasks using quantum mechanical systems. These tasks range from computational ones dealing with using quantum mechanics and quantum entanglement to solve computational problems, a subject which is of great theoretical and experimental interest, and information theoretical tasks such as cryptography, coding, data hiding etc. Entanglement, as a standalone subject is also researched in great detail, particularly the theory of entanglement monotones is a focus of interest.  

The main subject of this thesis is the state discrimination problem in quantum mechanics. Quantum state discrimination is a very fundamental problem in quantum theory and it is a great area to explore the very nature of quantum mechanics. The problem is very simple to state: given a quantum system in a known ensemble of quantum states, can the quantum state of the system be determined? It turns out that the answer is not a simple yes or no and the problem is very intimately related with the theory of quantum measurements. The applications of quantum state discrimination are wide, during any process where the state of the system must be determined by an observer, a quantum state discrimination scheme must be implemented. These situations arise particularly in quantum cryptography where the reciever of the message or the key must distinguish the state of the quantum systems which in fact are the carriers of information themselves, and in quantum computation, where the observer might be in a situation in which in order to learn the result of the computation, she must determine the state of the quantum system on which the result  of the computation is encoded. 

In the following chapters, many aspects of quantum state discrimination will be explored in detail, different strategies will be reviewed and the connection and the usefulness of the theory of majorization in characterizing quantum state discrimination problems will be presented. The possibility of discrimination of entangled quantum states while preserving their entanglement will also be discussed and the processes will be characterized using majorization and some partial results about the subject will be given.

In the opening chapter, some preliminary information about quantum mechanics, quantum measurements and entanglement will be presented. The review will be short and dense and its only purpose is to help the already familiar reader remember some key concepts.

In chapter 3, the quantum state discrimination problem will be presented and various aspects of it will be worked out. Several different cases of the problem such as monopartite and multipartite state discrimination will be investigated and different cases of the problem will be reviewed. Also, the theory of majorization and its connections with quantum information theory, more specifically its connections with quantum state discrimination will be explored.

In chapter 4, the problem of discrimination of entangled states with remaining entanglement will be presented and some special cases of the problem will be characterized using majorization and some partial results about the problem will be presented. Also the validity of the results will be discussed.

Lastly, in chapter 5, a short and concise review of the subjects covered in this work will be given and the results found in chapter 4 will be discussed and compared with the existing work in literature if applicable and the study will be concluded with some final remarks. Some fundamental results of probability theory relevant to the work are collected as an appendix.

\section{Preliminaries}
\subsection{Density Operators}
Central to the discussion of quantum mechanics and state discrimination problems is the mathematical tool of density matrices. The need for density matrices is straightforward, on many occasions the system in question is not in a definite quantum system but rather it is a statistical ensemble of different quantum states. An electron, for example might have its spin along the $+z$ direction with probability $p_{1}$ and along the $-z$ direction with probability $p_{2}$. The state of the electron cannot be represented as \begin{equation}
\ket{\psi} = \sqrt{p_{1}}\ket{0} + \sqrt{p_{2}}\ket{1} ,
\end{equation} 
where $\ket{0}$ is identified with the spin state along the $+z$ direction and $\ket{1}$ with the spin state along the $-z$ direction. The quantum state $\ket{\psi}$ in eq. (1.1) is in a coherent superposition of $\ket{0}$ and $\ket{1}$. The electron's spin is neither in the $+z$ direction nor in the $-z$ direction but, when one measures the spin along the $z$ direction, the electron will be in the state $\ket{0}$ with probability $p_{0}$ and it will be in the state $\ket{1}$ with probability $p_{1}$. But, the electron is not in a statistical mixture of states, the electron is in the state $|\psi\rangle$ with probability 1. In other words, the electron is in a \textquotedblleft pure\textquotedblright $\,$state. Suppose now the spin along the $z$ direction is measured. The outcomes are $+1$ and $-1$ with probabilities $p_{1}$ and $p_{2}$ respectively. If an observer performs the measurement but does not record the outcome, the quantum state of the electron must be represented by an ensemble 
\begin{equation}
\mathcal{E}=\{ (p_{1},\ket{0});(p_{2},\ket{1}) \}.
\end{equation}
More generally, an ensemble of pure states is a set of pairs 
\begin{equation}
\mathcal{E}=\{ (p_{i},\ket{\phi_{i}}) \}_{i=1}^{n} \, \, \text{and} \, \, \sum p_{i}=1.
\end{equation} with the meaning that the system is in the state $|\phi_{i}\rangle$ with probability $p_{i}$. A convenient way to express ensembles is to use the density operator defined as 
\begin{equation}
\rho = \sum_{i=1}^{n} p_{i}\dyad{\phi_{i}}, \, \braket{\phi_{i}}{\phi_{i}}=1.
\end{equation}
If the states $|\phi_{i}\rangle$ are re-expressed as a superposition of some orthogonal basis kets, \begin{equation}
\rho=\sum_{i,n,m} p_{i}c_{ni}c^{*}_{mi}\op{n}{m}, \, \ket{\phi_{i}}= \sum_{n} c_{ni}\ket{n} \, , \, \braket{n}{m} = \delta_{nm}.
\end{equation}
Since $p_{i}$ are real numbers, it is trivial to see that density operators are hermitian.

The expectation value of observables are also easy to calculate
\begin{equation}
\begin{split}
\langle A \rangle &= \sum_{i} p_{i} \langle A \rangle_{i}, \\ &=\sum_{i} p_{i} \expval{A}{\phi_{i}}, \\ &= \Tr(\sum_{i} A p_{i}\dyad{\phi_{i}}), \\ &= \Tr(\rho A).
\end{split}
\end{equation}
where the identity \begin{equation}
\braket{\psi}{\phi} = \Tr(\op{\phi}{\psi})
\end{equation}
is used. The trace of the density operator gives the normalization condition
\begin{equation}
\begin{split}
\Tr(\rho) &= \Tr(\sum_{i}p_{i}\dyad{\phi_{i}}), \\ &= \sum_{i}p_{i}\Tr(\dyad{\phi_{i}}), \\ &= \sum_{i}p_{i} \braket{\phi_{i}}, \\ &= \sum_{i} p_{i},\\ &=1
\end{split}
\end{equation}
since $\braket{\phi_{i}}{\phi_{i}}=1.$ The density operator is also a positive semidefinite operator, $\rho \geq 0$. This is quite easy to prove, for any arbitrary state ket $\ket{\psi}$
\begin{equation}
\begin{split}
\expval{\rho}{\psi} &= \sum_{i}p_{i} \braket{\psi}{\phi_{i}} \braket{\phi_{i}}{\psi}, \\ &= \sum_{i}p_{i} \abs{\braket{\psi}{\phi_{i}}} ^{2}, \\ &\geq 0.
\end{split}
\end{equation}
Thus, any positive semidefinite operator with trace equal to one represents an ensemble of pure quantum states. The correspondence between density operators and ensembles is not one-to-one, it is one-to-many. The same density operator might represent different ensembles with different pure quantum states. To see this, looking at the spectral decomposition of the density operator is enough;
\begin{equation}
\begin{split}
\rho &= \sum_{i}p_{i}\dyad{\phi_{i}}, \\ &= \sum_{i}\omega_{i}\dyad{i} \, \, \text{where} \, \, \braket{i}{j} = \delta_{ij} \, \, \text{and} \, \, \rho \ket{i} = \omega_{i} \ket{i} .
\end{split}
\end{equation}
Owing to the freedom of choosing the eigenvectors, different ensembles can be represented by the same density operator. 
The evolution of density matrices under unitary transformations is also easy to construct. $\text{Let} \, \, \ket{\phi_{i}^{\prime}} = U\ket{\phi_{i}},$
\begin{equation}
\begin{split}
\rho^{\prime} &= \sum_{i}p_{i}\dyad{\phi_{i}^{\prime}}, \\ &= \sum_{i}p_{i}U\dyad{\phi_{i}}U^{\dagger}, \\ &= U\rho U^{\dagger}.
\end{split}
\end{equation}
Density operators are also a convenient way to describe composite systems; systems whose quantum states live in a Hilbert space which is a direct product of two or more Hilbert spaces. The use of density matrices to describe composite systems and subsystems is through the reduced density matrix formalism. Suppose the system to be described lives in the Hilbert space $\mathcal{H} = \mathcal{H}_{A} \otimes \mathcal{H}_{B}$. The density operator $\rho_{AB}$ is a linear operator on $\mathcal{H}_{A} \otimes \mathcal{H}_{B}$. The operator $\rho_{AB}$ describes the ensemble of pure quantum states that are elements of the composite Hilbert space. To reach the description of a subsystem of the composite system, the reduced density matrix is used and it is defined through the partial trace operation
\begin{equation}
\rho_{A} = \Tr_{B} (\rho_{AB}),
\end{equation}
and the map of operators $\Tr_{B}$ is defined as
\begin{equation}
\begin{split}
\Tr_{B}(|\alpha_{1}\rangle_{A} \langle \alpha_{2}| \otimes |\beta_{1} \rangle_{B} \langle \beta_{2}|) &= |\alpha_{1}\rangle_{A} \langle \alpha_{2}| \Tr(|\beta_{1} \rangle_{B} \langle \beta_{2}|), \\ &= |\alpha_{1}\rangle_{A} \langle \alpha_{2}|(_{B}\langle \beta_{2}|\beta_{1}\rangle_{B}).
\end{split}
\end{equation}
For quantum states which can be expressed as $\rho_{AB} = \rho_{A}\otimes \rho_{B}$, the reduced density matrices are simply $\rho_{A}$ and $\rho_{B}$ due to the trace condition.

All mixed states represented by a density operator can be \textquotedblleft purified\textquotedblright $\,$ with the help of an ancilla system. The system is imagined to be entangled with an ancilla system, the ancilla might be the environment that the system is interacting with. An ancilla in the context of quantum information science is a system which is discarded at the end of the computation, it is merely a mathematical tool of no physical significance. Purification of a density operator is very straightforward, if $\rho=\sum_{i}p_{i}\dyad{\phi_{i}}, \, \text{then} \, \ket{\psi}_{AB}=\sum_{i}\sqrt{p_{i}}\ket{\phi_{i}}_{A} \otimes \ket{i}_{B}$ is a purification of $\rho$ meaning that the state of the quantum system in question can be expressed as a pure state of the system and the ancilla.

The density operator $\rho$ can be obtained by taking the partial trace over $B$
\begin{equation}
\begin{split}
\Tr_{B}(\dyad{\psi}) &= \sum_{i,j}\sqrt{p_{i}p_{j}}\op{\phi_{i}}{\phi_{j}}(\bra{i}\ket{j}), \\ &= \sum_{i,j}\sqrt{p_{i}p_{j}}\op{\phi_{i}}{\phi_{j}}\delta_{ij}. \\ &= \sum_{i}p_{i}\dyad{\phi_{i}} = \rho.
\end{split}
\end{equation}
Note that, due to the freedom of choosing the ancilla, purifications are not unique, but they are related by a unitary transformation on the ancilla.

\subsection{Quantum Measurements}
One of the main ways that quantum mechanics differ from its classical counterpart is the effect of measurements performed on quantum systems. Quantum mechanics is a perfectly deterministic theory unless measurements are included, states evolve unitarily and knowing the state of the system at time $t_{1}$ and the Hamiltonian $H$ of the system gives the description of the state of the system at a later time $t_{2}$ through the use of the unitary time evolution operator. But, measurement breaks the determinism, measurement outcomes in quantum mechanics is indeterministic, they give rise to a probability distribution of outcomes. To mathematically describe the process of measurement, the measurement formalism is introduced.

Quantum measurements are described by a set of measurement operators $\mathcal{M}=\{ M_{m} \}$ which are linear operators defined on the Hilbert space of the system to be measured. The index $m$ is a label for possible outcomes. 

Suppose that the quantum state of the system immediately before measurement is represented by the ket $\ket{\psi}$. The probability to obtain the outcome $m$ is given by
\begin{equation}
p_{m}=\expval{M_{m}^{\dagger} M_{m}}{\psi}
\end{equation}
and the quantum state of the system after measurement, corresponding to the $m^{th}$ outcome becomes
\begin{equation}
\ket{\phi_{m}} = \frac{M_{m}\ket{\psi}}{\sqrt{\expval{M_{m}^{\dagger} M_{m}}{\psi}}} .
\end{equation}
There is one equality that the measurement operators must satisfy which arises from the fact that probabilities must sum up to one;
\begin{equation}
1=\sum_{m}p_{m}=\sum_{m}\expval{M_{m}^{\dagger} M_{m}}{\psi}.
\end{equation}
Since the above condition must be satisfied for all $\ket{\psi}$, it is equivalent to
\begin{equation}
\sum_{m}M_{m}^{\dagger}M_{m} = \mathds{1}
\end{equation}
The completeness relation can be satisfied in many ways, the simplest one is to use a complete set of orthogonal projectors $P_{m}$ which satisfy
\begin{equation}
\sum_{m}P_{m} = \mathds{1} \, \, \, \text{and} \, \, \, P_{m}P_{n}=\delta_{mn}P_{n} .
\end{equation}
The second condition is simply the idempotence and orthogonality relations combined. Any hermitian operator, in other words an observable can be decomposed into its eigenvalues and projectors $M=\sum_{m}mP_{m}$ and the probability that the outcome $m$ is obtained in a measurement is simply 
\begin{equation}
p_{m} = \expval{P_{m}}{\psi} 
\end{equation} 
and the resultant state corresponding to the $m^{th}$ outcome is
\begin{equation}
\ket{\phi_{m}}= \frac{P_{m}\ket{\psi}}{\sqrt{\expval{P_{m}}{\psi}}}.
\end{equation}
These type of measurements are called von Neumann or projective measurements and they are a special case of a broader class of measurements. The effect of measurements on density matrices is also important and easy to formulate. Measurements map density operators to density operators. 

Consider an ensemble $\mathcal{E}=\{(p_{i},\ket{\phi_{i}})\}_{i=1}^{n}$.   Defining the conditional probability $p(m|i)$ as the probability that outcome $m$ is obtained given that the state is $\ket{\phi_{i}}$
\begin{equation}
\begin{split}
p(m|i) &=\expval{P_{m}}{\phi_{i}}, \\ &= \Tr(P_{m}\dyad{\phi_{i}}).
\end{split}
\end{equation}
Using the law of total probability (see Appendix), 
\begin{equation}
\begin{split}
p_{m} &= \sum_{i}p(m|i)p_{i}, \\ &= \sum_{i}p_{i} \Tr(P_{m}\dyad{\phi_{i}}), \\ &= \Tr(P_{m}\rho)
\end{split}
\end{equation}
is obtained, which is the probability of obtaining the outcome $m$ after measurement. Noting that the post measurement states are given as
\begin{equation}
\ket{\phi_{i}^{m}} = \frac{P_{m}\ket{\phi_{i}}}{\sqrt{\expval{P_{m}}{\phi_{i}}}}.
\end{equation}
The posterior density operator corresponding to the outcome $m$ is found to be;
\begin{equation}
\begin{split}
\rho_{m} &=\sum_{i}p(i|m)\dyad{\phi_{i}^{m}}, \\ &= \sum_{i}p(i|m)\frac{P_{m}\dyad{\phi_{i}}P_{m}}{\expval{P_{m}}{\phi_{i}}}, \\ &= \sum_{i} \frac{p(m|i)p_{i}}{p_{m}} \frac{P_{m}\dyad{\phi_{i}}P_{m}}{p(m|i)}, \\ &= \frac{1}{p_{m}}\sum_{i}p_{i}P_{m}\dyad{\phi_{i}}P_{m}, \\ \rho_{m} &= \frac{P_{m}\rho P_{m}}{\Tr(P_{m}\rho)}
\end{split}
\end{equation}
where the Bayes rule, expressed as
\begin{equation}
p(i|m)=\frac{p(m|i)p_{i}}{p_{m}}
\end{equation}
is used, giving the probability that the state is $\ket{\phi_{i}} $ given that the outcome is $m$. (Refer to the Appendix for a brief review of probability theory.) 

Sometimes, in the discussion of quantum information theory, the concept of forgetful measurements gain importance. The idea is that, after performing the measurement, the observer does not record the outcome. The resultant quantum state is then a density operator formed by the corresponding outcomes and the associated probabilities, Suppose the posterior state corresponding to outcome $m$ are called as $\rho_{m}$. The outcome $m$ has a probability of $p_{m}$. The density operator after a forgetful measurement can be expressed as;
\begin{equation}
\begin{split}
\rho^{\prime} &= \sum_{m}p_{m}\rho_{m}, \\ &= \sum_{m}\Tr(P_{m}\rho) \frac{P_{m}\rho P_{m}}{Tr(P_{m}\rho)}, \\ \rho^{\prime} &= \sum_{m}P_{m}\rho P_{m}.
\end{split}
\end{equation}

The discussion above covers all aspects of the measurement formalism known as projective measurements or von Neumann measurements. However, more general measurement formalism can be constructed which is much more powerful than its projective counterpart. This more general is called the generalized measurement formalism, or when only the probabilities of obtaining the outcomes are concerned, it is sometimes called the POVM formalism where POVM stands for Positive Operator-Valued Measure \cite{nielsenbook}.

A measurement described by the operators $\mathcal{M}=\{M_{m}\}$ must always satisfy the rule that probabilities $p_{m}$ corresponding to the outcomes $m$ must satisfy $\sum_{m}p_{m} = 1$. In terms of the operators themselves, this can be expressed as,
\begin{equation}
E_{m} \equiv M_{m}^{\dagger}M_{m} \, \, , \, \, \sum_{m}E_{m} =\mathds{1}
\end{equation}
where the probability of obtaining $m$ is given by $p_{m} = \expval{E_{m}}{\psi}$. It is easy to see that the operators $E_{m}$ are positive operators since $p_{m} \geq 0$. Since any positive operator can be decomposed as $A = B^{\dagger}B$, the set of operators $E_{m}$ are sufficient to describe a measurement $\mathcal{M}$. The operators $E_{m}$ are known as the POVM elements associated with the outcome $m$ for a measurement $\mathcal{M}=\{M_{m}\}$. Choosing $E_{m}$ to be projectors $P_{m} = \dyad{\phi_{m}}$ is a special case where the number of different outcomes must be equal to or smaller than the dimension of the Hilbert space $\mathcal{H}$. For a POVM however, the number of outcomes $m$ can be bigger than $\dim \mathcal{H}$  
since the general measurement elements $E_{m}=M_{m}^{\dagger}M_{m}$ do not obey any orthogonality relation. This fact will be quite important when quantum state discrimination is discussed.

The effect of measurement on quantum states and density operators should also be stated. The results will be generalized from the projective case.

The posterior state after the measurement and the probability corresponding to outcome $m$ is simply
\begin{equation}
\ket{\phi_{m}}=\frac{M_{m}\ket{\psi}}{\sqrt{\expval{M_{m}^{\dagger}M_{m}}{\psi}}} \, \, , \, \, p_{m}=\expval{M_{m}^{\dagger}M_{m}}{\psi}
\end{equation}
For density operators, the expressions can be obtained by generalizing the results for projective measurements. The probability of obtaining outcome $m$ is given by
\begin{equation}
p_{m} = \Tr(M_{m}^{\dagger} M_{m} \rho)
\end{equation}
and the density operator corresponding to outcome $m$ is given by
\begin{equation}
\rho_{m}=\frac{M_{m} \rho M_{m}^{\dagger}}{\Tr(M_{m}^{\dagger}M_{m}\rho)}
\end{equation}
and, for a forgetful measurement in which the outcome is not recorded by the observer,
\begin{equation}
\rho^{\prime}=\sum_{m}M_{m}\rho M_{m}^{\dagger}.
\end{equation}
The measurement formalism in quantum mechanics can be extended to measurements done on composite systems. In that case, the measurement operators act only on the designated subspace of the total Hilbert space $\mathcal{H}=\mathcal{H}_{A} \otimes \mathcal{H}_{B}$. A measurement $\mathcal{M}={M_{m}}$ is separable if all the operators $M_{m}$ are of the form
\begin{equation}
M_{m}=F^{A}_{m} \otimes F_{m}^{B} 
\end{equation}
where the superscripts $A$ and $B$ denote the subspaces that the operators $F_{m}$ act on. When dealing with composite systems, an operator $M_{m}^{A}$ is a shorthand for $M_{m}^{A} \otimes \mathds{1}^{B}$. Certain types of measurements in which the result that one party obtains after measuring her subsystem is conditional on the result of the other party are called LOCC protocols where LOCC stands for Local Operations and Classical Communication which are a subset of separable measurements but mathematically cannot in general be expressed as easily and cleanly as separable measurements. LOCC protocols will be explained in detail in chapter 3. 

\subsection{Entanglement}
Quantum entanglement is the most interesting and baffling aspect of quantum mechanics. Since the foundation of quantum mechanics in its modern form, much debate has been centered on the subject of entanglement and there is an impressive amount of literature devoted to the subject. Stated in words, quantum entanglement is the phenomena in which the measurement results obtained from the different subsystems are correlated in a way that is much stronger than classical correlations. Although the statement above might seem imprecise and confusing, the mathematical framework of quantum entanglement is well founded and rigorous. The framework will be explored in detail below. 

Composite quantum states might come in various forms, the simplest form is what is called as the "product" form, meaning that if a state $\ket{\psi}_{AB}$ can be expressed as 
\begin{equation}
\ket{\psi}_{AB} = \ket{\phi}_{A} \otimes \ket{\varphi}_{B}
\end{equation}
then, $\ket{\psi}_{AB}$ is a product state. A general definition can be made, any pure quantum state that cannot be expressed in product form is an entangled state.

An entangled state lives in the direct product space of the respective Hilbert spaces of the subsystems. For a bipartite (two subsystems) state, the most general expression for the state can be written as
\begin{equation}
\ket{\psi}=\sum_{i,j}c_{ij}\ket{i}_{A}\ket{j}_{B}
\end{equation}
where the $\otimes$ symbol has been omitted as the meaning is clear. The set of vectors $\{\ket{i}_{A}\ket{j}_{B}\}_{i=1,j=1}^{n,m}$ is an orthonormal basis for the Hilbert space $\mathcal{H}_{A} \otimes \mathcal{H}_{B}$. 

The concept of reduced density matrices is widely used when dealing with entangled states and the rank of the reduced density operators is directly related with the entanglement in the following way: a quantum state for which the reduced density operators rank is greater than 1 is an entangled state.

To better understand the statement, the Schmidt decomposition of entangled states must be understood. The Schmidt decomposition is a procedure after which the state is expressed as a superposition of pairwise matched vectors of two orthonormal bases of the Hilbert spaces of the subsystems. Any bipartite quantum state can be expressed in Schmidt form;
\begin{equation}
\ket{\psi}=\sum_{i}\lambda_{i}\ket{i}_{A}\ket{i}_{B}
\end{equation}
where $\ket{i}_{A} \, \text{and} \ket{i}_{B}$ are complete orthonormal bases for their respective Hilbert spaces and $\lambda_{i}$ are known as the Schmidt coefficients and they are also the square roots of the eigenvalues of the reduced density matrices. A quick note is in order here, $\ket{i}_{A}$ and $\ket{i}_{B}$ might represent states of different physical nature; and $i$ here is just a label. If a quantum state has only one Schmidt coefficient in its Schmidt decomposition, than it is a product state and since the Schmidt coefficients are square roots of the eigenvalues of the reduced density matrices, a state having a rank 1 reduced density matrix is a product state.

The Schmidt decomposition is formally achieved through the singular value decomposition procedure: if $c$ is an arbitrary (possibly rectangular) matrix, than there are unitary matrices $u$ and $v$ and the possibly rectangular, diagonal matrix $d$ with diagonal elements being the singular values, the singular value decomposition is 
\begin{equation}
c=udv \, \, \text{or} \, \, c_{ij}=\sum_{k}u_{ik}d_{kk}v_{kj}.
\end{equation}

The proof of the Schmidt decomposition is as follows;
\begin{equation}
\begin{split}
\ket{\psi} &= \sum_{i,j}c_{ij}\ket{a_i}\ket{b_j}, \\ &= \sum_{i,j,k}u_{ik}d_{kk}v_{kj}\ket{a_i}\ket{b_j} \, \, \text{after singular value decomposition} \\ &= \sum_{k}\lambda_{k}\ket{k}_{A}\ket{k}_{B}
\end{split}
\end{equation}
after defining $\sum_{k}u_{ik}\ket{a_i} = \ket{k}_{A} \, \, , \, \, \sum_{j}v_{kj}\ket{b_j} = \ket{k}_{B} \, \, \text{and} \, \, d_{kk}=\lambda_{k}.$

The density operator associated with $\ket{\psi}$ is $\rho_{AB}=\ket{\psi}\bra{\psi}$. Partial tracing over B gives;
\begin{equation}
\begin{split}
\Tr_{B}\rho_{AB} &= \sum_{i,j}\lambda_{i}\lambda_{j}\ket{i}_{A}\bra{j}(\bra{i}\ket{j}), \\ &= \sum_{i}\lambda_{i}^{2}\dyad{i}.
\end{split}
\end{equation}
Note that the expression on the last line of eq. (2.39) is the spectral decomposition for $\rho_{A}$ with $\lambda_{i}^{2}$ as the eigenvalues and the rank of the reduced density operator is the number of Schmidt coefficients. A rank 1 reduced density operator will correspond to a product state since the state will be of the form $\ket{\psi}=\ket{k}_{A}\otimes \ket{k}_{B}$. It is clear from here that the Schmidt coefficients for a bipartite state is the square roots of the eigenvalues of the reduced density matrices. Note that the above result doesn't change when the trace is taken over $A$, the eigenvalues of both reduced density operators are the same.  In general the Schmidt decomposition can only be applied to bipartite states, it fails for states with more than two subsystems bar a few special cases like the GHZ state \cite{GHZ}. 

The amount of entanglement that a state possesses can be quantified by a few methods. These quantifiers are called entanglement monotones and they are required not to increase in any way during LOCC protocols. The study of entanglement monotones is a rich and broad subject in itself so not much detail is going to be given here but two of those monotones; entanglement entropy and concurrence are going to be explained in a quick fashion. 

Entanglement entropy is defined as 
\begin{equation}
\begin{split}
\mathcal{E}(\rho_{AB}) &=-\Tr (\rho_{A}\log_{2} \rho_{A}), \\ &= -\sum_{i}p_{i}\log_{2} p_{i}
\end{split}
\end{equation}
where $\rho_{A},\rho_{B}$ are the reduced density matrices of the quantum state and $p_{i}$ are the eigenvalues of the reduced density matrix. The unit of entanglement is usually called ebit and a bipartite state can have at most 1 ebit of entanglement. For two level systems, or qubits, the maximally entangled states are also called as the Bell states and in the computational basis $\{\ket{0},\ket{1}\}$, they can be expressed as follows,
\begin{equation}
\begin{split}
\ket{\psi^{+}} &= \frac{1}{\sqrt{2}}\ket{00}+\frac{1}{\sqrt{2}}\ket{11}, \\
\ket{\psi^{-}} &= \frac{1}{\sqrt{2}}\ket{00}-\frac{1}{\sqrt{2}}\ket{11}, \\
\ket{\phi^{+}} &= \frac{1}{\sqrt{2}}\ket{01} + \frac{1}{\sqrt{2}}\ket{10}, \\
\ket{\phi^{-}} &= \frac{1}{\sqrt{2}}\ket{01} - \frac{1}{\sqrt{2}}\ket{10}.
\end{split}
\end{equation}
Note that the reduced density matrix for each of these states is $\rho_{A} = \frac{1}{2}\mathds{1}$, the maximally mixed ensemble. Hence the following statement can be made, a state is maximally entangled if its reduced density matrix corresponding to one of the systems is $\frac{1}{N}\mathds{1}$ where $N$ is the dimension of the corresponding Hilbert space. 

Another easy to calculate entanglement monotone is the concurrence defined for a mixed state of two qubits (2 level systems) as;
\begin{equation}
\mathcal{C}(\rho) = \max(0,\lambda_{1},\lambda_{2},\lambda_{3},\lambda_{4})
\end{equation}
where $\lambda_{i}$ are the eigenvalues of the hermitian matrix $\sigma=\sqrt{\sqrt{\rho}\tilde{\rho}\sqrt{\rho}}$ and $\tilde{\rho}$ is defined as $\tilde{\rho}=(\sigma_{y} \otimes \sigma_{y})\rho^{\star}(\sigma_{y} \otimes \sigma_{y})$. For pure states which can be written in the Bell form $\ket{\psi}=a\ket{00}+b\ket{11}$, the concurrence reduces to;
\begin{equation}
\mathcal{C}(\psi) = 2\abs{a}\abs{b}.
\end{equation}

There are various other quantifiers of the amount of entanglement in a quantum state, in fact, any concave function of the squares of the Schmidt coefficients of a state is an entanglement monotone, but they are not useful for the purposes of this work. For the discussion here, concurrence and entanglement entropy will be sufficient.

\section{Quantum State Discrimination}

Quantum state discrimination is a fundamental problem in the theory of quantum information and much research has been devoted to this subject over the years. The problem was first noticed by Helstrom \cite{helstromdisc} in 1976 and Holevo \cite{holevodisc} in 1979. Later in 1987-1988, Ivanovic \cite{Ivan}, Peres \cite{peres} and Dieks \cite{dieks} showed that unambiguous discrimination of linearly independent states is possible and Chefles \cite{chefles} in 1998 generalized the problem to quantum state separation. Nowadays, the research is more focused on LOCC discrimination of multipartite states, started by Walgate \cite{walgate} and his colleagues. In itself, quantum state discrimination is a simple, easy to state problem; given an ensemble of quantum states, can the observer determine the state of the system? It turns out that the solution is trivial if the states to be discriminated are orthogonal to each other but if they are linearly independent but not orthogonal, the solution is not that easy and in most cases, due to the very nature of quantum mechanics, the only possibility is to discriminate with some error. It also turns out that for general cases the problem cannot be analytically solved, the analytical solutions only exist for a few special cases. 

There are many approaches developed to solve the problem of state discrimination such as minimum error discrimination, unambiguous discrimination and others but all the methods carry an inherent probability of failure in different aspects. It turns out that the problem is actually a linear optimization problem, the task is to minimize the probability of error for a given ensemble. 

In this chapter, the problem of state discrimination will be formally defined and it will be shown that why perfect (error less)  discrimination is not possible for non-orthogonal states and then review the minimum error discrimination and unambiguous discrimination strategies for monopartite states. The discussion will than move on to the discrimination of bipartite, possibly entangled, orthogonal states under LOCC. There are many more approaches for solving the stated problem but they are not relevant for the purposes of this study. The body of the chapter will be based on a tutorial review by J. A. Bergou\cite{review}.

\subsection{Statement of the Problem}
In a formal mathematical setting, the quantum state discrimination problem is easy to describe: given an ensemble $\mathcal{E}=\{(p_{i},\ket{\psi_{i}})\}_{i=1}^{n}$, can the state of the system be determined? The answer is easy if $\ket{\psi_{i}}$ are mutually orthogonal, it is possible to perfectly discriminate between the states without any error. But, if they are not mutually orthogonal but linearly independent, than it is still possible to determine the state of the system with some probability of error. In the most general setting, the states $\ket{\psi_{i}}$ are upgraded to density operators $\rho_{i}$. 

First, the notion of perfect discrimination will be defined and the cases when it is possible will be explained. For an ensemble $\mathcal{E}=\{(p_{i},\ket{\psi_{i}})\}$ and the measurement $M=\{M_{j}\}$ with POVM elements $E_{i}=M_{i}^{\dagger}M_{i}$, perfect discrimination is possible if
\begin{equation}
M_{j}\ket{\psi_{i}}=0 \, \, \text{when} \, \, i \neq j .
\end{equation}
In other words, when the measurement is performed, for the state $\ket{\psi_{i}}$ if all the measurement outcomes other than the $i^{th}$ one occur with 0 probability, then it is concluded that the state was originally $\ket{\psi_{i}}$. If $\ket{\psi_{i}}$ are mutually orthogonal, a  complete projective measurement with elements $P_{j}$ can be constructed such that $P_{j}\ket{\psi_{i}}=0$  when $i \neq j$. A quick note, if the given states $\ket{\psi_{i}}$ do not form a complete orthogonal set, the missing kets can be added to the ensemble with 0 probability and add the projector $P_{0} = \mathds{1} -\sum_{i}P_{i}$ to the list of projectors and construct a complete projective measurement. Since the added states occur with zero probability, the conditional probabilities of the corresponding outcomes are 0 enabling us to express that outcome with a single POVM element. Thus, if the given states are orthogonal, perfect discrimination using projectors is always possible. 
Showing that perfect discrimination is not possible if the states are not orthogonal is also easy. Suppose that the ensemble in question consists of two states $\ket{\psi_{1}}$ and $\ket{\psi_{2}}$, not necessarily orthogonal and two operators, $M_{1}$ and $M_{2}$ are constructed to discriminate these states  obeying $M_{1}+M_{2}=\mathds{1}$. For perfect discrimination, the following must hold
\begin{equation}
\begin{split}
M_{1}\ket{\psi_{2}} &= 0, \\
M_{2}\ket{\psi_{1}} &= 0.
\end{split}
\end{equation}
The conditional probabilities are defined as 
\begin{equation}
\begin{split}
p(1|1) &=\expval{E_{1}}{\psi_{1}}, \\
p(2|2) &=\expval{E_{2}}{\psi_{2}}, \\
\end{split}
\end{equation}
and the others are 0 due to orthogonality. 
If the relation $M_{1}+M_{2}=\mathds{1}$is multiplied  with $\bra{\psi_{1}}$ from the left and with $\ket{\psi_{2}}$ from the right, 
\begin{equation}
\bra{\psi_{1}} E_{1}\ket{\psi_{2}} + \bra{\psi_{1}} E_{2}\ket{\psi_{2}} = \braket{\psi_{1}}{\psi_{2}}.
\end{equation}
Taking into account eq.(3.2), it is seen that the above expression must be zero, which is only possible for orthogonal states. Thus, for perfect discrimination to be possible, the states to be discriminated should be orthogonal to each other. Below, the two main strategies to follow when the states are not orthogonal are going to be explained. 

\subsection{Minimum Error Discrimination}
The minimum error strategy was first coined by Helstrom \cite{helstromdisc} in 1976. The minimum error strategy aims to eliminate inconclusive results but permits errors in the measurement scheme. The problem than transforms into a linear optimization problem, finding the measurement operators that minimize the probability of error. 
As in all state discrimination problems, an ensemble of states $\mathcal{E}=\{(p_{i},\rho_{i})\}_{i=1}^{n}$ is given where each state $\rho_{i}$ occurs with prior probability $p_{i}$. A suitable measurement $\mathcal{M}=\{M_{j}\}$ where $\sum_{j}M^{\dagger}_{j}M_{j}=\mathds{1}$ with $E_{i}=M^{\dagger}_{i}M_{i}$ being the POVM elements, is used to discriminate between the states. The measurement operators are defined such that if it has been prepared in the state $\rho$ where $\rho=\sum_{i}p_{i}\rho_{i}$, the probability to conclude that the system is in the state $\rho_{i}$ is $\Tr(E_{i}\rho)$. The number of the measurement operators is also equal to the number of states to be discriminated since the measurement needs to be exhaustive, meaning that the measurement should cover all the outcomes and should not give inconclusive results.

The overall probability of error, where the error is the probability that the state is concluded to be $\rho_{j}$ but it was in fact $\rho_{i}$ where $i\neq j$ can be simply defined using the fact that probabilities add up to one.
\begin{equation}
p_{er}=1-\sum_{i}p_{i}\Tr(E_{i}\rho_{i}).
\end{equation}
The trace term $\Tr(E_{i}\rho_{i})$ is the probability of correctly identifying the state $\rho_{i}$ and the $p_{i}$ are the prior probabilities of the states, hence the second term in the right hand side of eq.(3.5) is the probability that the result is correct. The best protocol is achieved when the probability of error is minimized by finding the optimal set of operators $M_{m}$. This, however, is no easy task and there is not any general solution for an arbitrary number of states. But, an analytical solution for 2 states can be found. 

Consider an ensemble $\mathcal{E}=\{(p_{i},\rho_{i})\}_{i=1}^{2}$ and the POVM elements $E_{1}$ and $E_{2}$ which satisfy $E_{1}+E_{2}=\mathds{1}$. Using eq. (3.5), the error probability can be found as
\begin{equation}
\begin{split}
p_{er} &= 1-p_{1}\Tr(E_{1}\rho_{1}) - p_{2}\Tr(E_{2}\rho_{2}), \\ &= p_{1}\Tr(E_{2}\rho_{1}) + p_{2}\Tr(E_{1}\rho_{2}).
\end{split}
\end{equation}
The second line is also equal to the total probability of an erroneous detection since the trace terms are the probability of false detection events. The error probability can be cast in another form with the help of a new hermitian operator defined as
\begin{equation}
\begin{split}
\Lambda &= p_{2}\rho_{2}-p_{1}\rho_{1}, \\ &= \sum_{j}^{D}\lambda_{k}\dyad{k},
\end{split}
\end{equation}
where the second line in eq. (3.7) is the spectral decomposition of $\Lambda$.
Inserting $p_{1}\rho_{1}=\Lambda-p_{2}\rho_{2}$ into the second line of eq. (2.6),
\begin{equation}
\begin{split}
p_{er} &= \Tr[(p_{2}\rho_{2}-\Lambda )E_{2}]+p_{2}\Tr(\rho_{2}E_{1}), \\  &= p_{2} - \Tr[\rho_{2}(E_{1}+E_{2})]- \Tr (\Lambda  E_{2}), \\ p_{er} &= p_{2} - \Tr (\Lambda E_{2})
\end{split}
\end{equation}
where the facts that $E_{1}+E_{2}=\mathds{1}$ and $\Tr \rho = 1$ were used. Note that $p_{er} = p_{1} + \Tr(\Lambda E_{1})$ if $p_{2}\rho_{2} = \Lambda + p_{1}\rho_{1}$ is inserted into eq. (3.6).

Since $\Lambda$ is hermitian, its eigenvalues are all real, hence its eigenvalues can be classified into 3 subgroups without loss of any generality 
\begin{equation}
\begin{split}
\lambda_{k} &< 0 \, \, \text{for} \, \, 1 \leq k < k_{-}, \\ \lambda_{k} &= 0 \, \, \text{for} \, \, k_{-} \leq k \leq k_{0}. \\ \lambda_{k} &> 0 \, \, \text{for} \, \, k_{0} < k \leq D .
\end{split}
\end{equation}
where $D$ is the dimension of the Hilbert space. Inserting the spectral decomposition of $\Lambda$ into the expression for the probability of error in eq. (3.8), the following expression is obtained after using $\braket{\alpha}{\beta} = \Tr(\op{\beta}{\alpha})$
\begin{equation}
\begin{split}
p_{er} &= p_{2} - \sum_{k=1}^{D} \Tr(\lambda_{k} \dyad{k} E_{2}), \\
p_{er} &= p_{2} - \sum_{k=1}^{D}\lambda_{k} \bra{k}E_{2}\ket{k} = p_{1} + \sum_{k=1}^{D} \lambda_{k} \bra{k}E_{1}\ket{k}.
\end{split}
\end{equation}
Since the expectation values involving the operators $E_{1}$ and $E_{2}$ are probabilities, the measurement operators hence the expectation value terms in eq. (3.10) are positive semi-definite by construction. To minimize the error expression in eq. (3.10), the following equalities must hold.
\begin{equation}
\begin{split}
\expval{E_{2}}{k} &= 0 \, \, \text{and} \, \, \expval{E_{1}}{k} = 1 \, \, \text{for} \, \, \lambda_{k}<0, \\ \expval{E_{2}}{k} &= 1 \, \, \text{and} \, \, \expval{E_{1}}{k} = 0 \, \, \text{for} \, \, \lambda_{k}>0.
\end{split}
\end{equation}
The minimizing conditions can also be understood in the following way, since the error probability is a linear function of the expectation values $\expval{E_{1}}{k}$ and $\expval{E_{2}}{k}$, the extrema happen at the end points of these functions. The true minima happen to be the case described above. These conditions allow the operators $E_{i}$ constructed as
\begin{equation}
E_{1}=\sum_{k=1}^{k_{-}-1}\dyad{k} \, \, \text{and} \, \, E_{2}=\sum_{k=k_{-}}^{D}\dyad{k}
\end{equation}
Note that the operator $E_{2}$ also includes the eigenkets corresponding to 0 eigenvalues but the 0 eigenvalues have no effect on the error probability whatsoever. The eigenkets corresponding to the eigenvalues 0 can also be included into $E_{1}$, meaning that the solution is not unique. To find the minimum error probability, the expressions for the operators $E_{1}$ and $E_{2}$ are inserted into the last line of eq. (3.8)
\begin{equation}
\begin{split}
p_{er} &= p_{2} - \Tr(\sum_{k=1}^{D}\lambda_{k}\dyad{k} \sum_{k=k_{-}}^{D}\dyad{k}), \, \lambda_{k} \geq 0 \, \, \text{for} \, \, k_{-} \leq k \leq D, \\ &= p_{2} - \sum_{k=k_{-}}^{D}\abs{\lambda_{k}}, \\ p_{er} &= p_{1} + \Tr(\sum_{k=1}^{D}\lambda_{k}\dyad{k} \sum_{k=1}^{k_{-}}\dyad{k}) , \,  \lambda_{k} < 0 \, \, \text{for} \, \, 1 \leq k < k_{-}, \\ p_{er} &= p_{1} - \sum_{k=1}^{k_{-}}\abs{\lambda_{k}}.
\end{split}
\end{equation}
Adding up the two alternative expressions found in eq. (3.13) and dividing by two, the final expression can be reached
\begin{equation}
\begin{split}
p_{er} &= \frac{1}{2}(1-\sum_{k}\abs{\lambda_{k}}), \\ &= \frac{1}{2}(1-\Tr\abs{\Lambda }),
\end{split}
\end{equation}
where $\abs{A}$ is defined as $\sqrt{A^{\dagger}A}$ and it is trivial to see that all its eigenvalues are the norms of the original eigenvalues.
Inserting the expression for $\Lambda$ from eq. (3.7), the lower bound for the error probability is found, also known as the Helstrom bound
\begin{equation}
p_{er} = \frac{1}{2}(1-\Tr\abs{p_{2} \rho_{2}- p_{1} \rho_{1}}).
\end{equation}
This equation can be cast in another form if the states to be discriminated are pure states $\ket{\psi_{1}}, \, \ket{\psi_{2}}$ and thus $\rho_{1} = \dyad{\psi_{1}}$ and $\rho_{2} = \dyad{\psi_{2}}$. In this case, the minimum attainable probability of error can be expressed as
\begin{equation}
p_{er} = \frac{1}{2}(1-({1-4p_{1}p_{2}\abs{\braket{\psi_{1}}{\psi_{2}}}^{2}})^{1/2}).
\end{equation}
To get this result, first note that when the states to be discriminated are pure, the hermitian operator $\Lambda$ has the form
\begin{equation}
 \Lambda = p_{2}\dyad{\psi_{2}}-p_{1}\dyad{\psi_{1}}. 
\end{equation}
Since any two state defines a 2 dimensional Hilbert space, without loss of any generality $\ket{\psi_{1}}$ and $\ket{\psi_{2}}$ can be expressed as superpositions of arbitrary basis kets $\{ \ket{0},\ket{1} \}$
\begin{equation}
\begin{split}
\ket{\psi_{1}}&=\cos \theta \ket{0} + \sin \theta \ket{1}, \\ \ket{\psi_{2}}&=\cos \theta \ket{0} - \sin \theta \ket{1}.
\end{split}
\end{equation}
Inserting the expressions for the states in eq. (3.18) into eq. (3.17)
\begin{equation}
\Lambda = \begin{pmatrix}
(p_{2}-p_{1})\cos ^{2} \theta & -(p_{2}+p_{1})\cos \theta \sin \theta \\ -(p_{2}+p_{1})\cos \theta \sin \theta & (p_{2}-p_{1})\sin ^{2} \theta
\end{pmatrix}.
\end{equation}
is obtained. The eigenvalues of this matrix can be calculated easily and they are 
\begin{equation}
\lambda_{\pm}=\frac{1}{2}(p_{2}-p_{1} \pm (1-4p_{1}p_{2}\abs{\braket{\psi_{1}}{\psi_{2}}}^{2})^{1/2})
\end{equation}
where $\abs{\braket{\psi_{1}}{\psi_{2}}}^{2}=\cos^{2}2\theta$
Now, note that in the minimum error expression $p_{er}=p_{2}-\Tr(\Lambda M_{2})$, the operator $M_{2}$ projects onto the positive eigenvalue eigenstates of $\Lambda$ by construction. Using the positive eigenvalue in eq. (3.20) and inserting it into the last line of eq. (3.8), the minimum error probability for 2 pure states is shown to be
\begin{equation}
p_{er} = \frac{1}{2}(1-({1-4p_{1}p_{2}\abs{\braket{\psi_{1}}{\psi_{2}}}^{2}})^{1/2})
\end{equation}
which is the same as in eq. (3.16). There is an interesting result when $\Lambda \geq 0$, in that case $M_{1} = 0$ and $M_{2}=\mathds{1}$, meaning that the minimum probability of error can always be attained by simply guessing that the state is $\rho_{2}$ or $\ket{\psi_{2}}$.

There are other select cases for which an analytical solution for a minimum error discrimination scheme is possible like symmetric states and mirror symmetric states but they will not be explored here. For general problems, the necessary and sufficient conditions for a POVM that realizes a minimum error discrimination to exist are known
\begin{equation}
\begin{split}
\sum_{i}p_{i}\rho_{i}E_{i} - p_{j}\rho_{j} &\leq 0, \, \, \forall j, \\ E_{i}(p_{i}\rho_{i} - p_{j}\rho_{j})E_{j} &= 0, \, \, \forall i,j,
\end{split}
\end{equation}
but these results are not going to be proven. A proof of these conditions can be found in \cite{generalproof}. The discussion above covers the general aspects and the idea of the minimum error discrimination strategy. 

\subsection{Unambiguous Discrimination}
The main difference between the unambiguous discrimination and the minimum error strategies are straightforward, given an ensemble $\mathcal{E}=\{(p_{i},\ket{\psi_{i}}\}$ in UQSD (Unambiguous Quantum State Discrimination) the observer is not allowed to make an erroneous decision, if she states that the state is $\ket{\psi_{1}}$, the state is $\ket{\psi_{1}}$ with probability 1, unlike the minimum error discrimination strategy. In mathematical terms, this idea can be expressed as
\begin{equation}
\expval{E_{j}}{\psi_{i}} = 0 \, \, \forall j \neq i, 
\end{equation}
where $E_{j}$ are the POVM elements. As it was shown at the beginning of this chapter in eq. (3.4), non-orthogonal states cannot be perfectly distinguished from each other. This means that the operators $M_{j}$ cannot satisfy the completeness condition $\sum_{j}M^{\dagger}_{j}M_{j}=\mathds{1}$. To satisfy the completeness requirement, other POVM elements $E_{F\lambda}=M^{\dagger}_{F\lambda}M_{F\lambda}$ are introduced, which are sometimes also known as the failure operator, and for the general case the number of failure operators can be more than 1. These operators correspond to the outcomes for which no conclusion can be drawn about the state of the system. The POVM elements $E_{j}$ are the operators to identify the states $\ket{\psi_{j}}$, such that $\expval{E_{j}}{\psi_{j}}=p(j|j)$ is the probability of successfully identifying the state $\ket{\psi_{j}}$ and $\sum_{\lambda}\expval{E_{F\lambda}}{\psi_{i}} = p^{F}_{i}$ is the probability of failing to identify the state $\ket{\psi_{i}}$. As usual the POVM elements must satisfy the following conditions
\begin{equation}
\begin{split}
\sum_{j}E_{j}+\sum_{\lambda}E_{F\lambda}&=\mathds{1}, \\ E_{j} &\geq 0, \, \forall j, \\ E_{F\lambda} &\geq 0, \, \forall \lambda.
\end{split}
\end{equation}
Finding the optimal unambiguous discrimination scheme for a given ensemble is again a linear optimization problem over the POVM elements achieving the lowest probability of failure for a given ensemble.
The general idea behind UQSD is pretty simple. For a given ensemble,  with the state set $\{\psi_{i}\} _{i=1}^{N}$, a reciprocal set $\{\psi_{i}^{\perp} \}_{i=1}^{N}$ can be constructed where $\braket{\psi_{i}^{\perp}}{\psi_{j}}=0, \, \, \text{if} \, \, i \neq j$ by construction. The general idea is to project the states to be discriminated onto the reciprocal states, since the expectation value of a reciprocal projector in the non-conjugate state is 0 for all the states other than the corresponding state in the original set, the detection is unambiguous but to satisfy the completeness condition of POVMs, operators corresponding to inconclusive results must be added to the measurement set. To understand the idea better, an analytical solution for two states will be presented.

Suppose the initial ensemble consists of two linearly independent but not necessarily orthogonal states $\mathcal{E}=\{(p_{1},\ket{\psi_{1}}),(p_{2},\ket{\psi_{2}}\}$. The reciprocal set of states is defined as, $\{\ket{\psi_{1}^{\perp}},\ket{\psi_{2}^{\perp}}\}$ where $\braket{\psi_{i}^{\perp}}{\psi_{j}}=0, \, \, \text{if} \, \, i \neq j$. The measurement operators corresponding to successful identification must satisfy
\begin{equation}
\expval{E_{i}}{\psi_{j}}=0, \, \, \forall i \neq j.
\end{equation}
Using the idea to project onto the conjugate pair, the operators $E_{1}$ and $E_{2}$ can be written as
\begin{equation}
\begin{split}
E_{1} &= c_{1}\dyad{\psi_{1}^{\perp}}, \\ E_{2} &= c_{2}\dyad{\psi_{2}^{\perp}}.
\end{split}
\end{equation}
It is easy to check that these operators satisfy the condition in eq. (3.25). That condition can also be used to solve for the coefficients $c_{1}$ and $c_{2}$ in eq. (3.26). It gives
\begin{equation}
\begin{split}
E_{1} &= \frac{\expval{E_{1}}{\psi_{1}}}{\abs{\braket{\psi_{1}}{\psi_{1}^{\perp}}}^{2}}\dyad{\psi_{1}^{\perp}}, \\ E_{2} &= \frac{\expval{E_{2}}{\psi_{2}}}{\abs{\braket{\psi_{2}}{\psi_{2}^{\perp}}}^{2}}\dyad{\psi_{2}^{\perp}}.
\end{split}
\end{equation}
These operators are positive semi-definite by construction but for the existence of a POVM that can realize the discrimination scheme, the operator corresponding to the inconclusive outcome $E_{F}=\mathds{1}-E_{1}-E_{2}$ must also be positive semi-definite. The positive semi-definiteness of the failure operator can be thought as the following inequality
\begin{equation}
E_{1}+E_{2} \leq \mathds{1}
\end{equation}
Since the set $\{\ket{\psi_{1}},\ket{\psi_{2}}\}$ defines a 2 dimensional Hilbert space, any arbitrary state $\ket{\Psi}$ can be expanded as a superposition of the states $\ket{\psi_{1}},\ket{\psi_{2}}$ as 
\begin{equation}
\ket{\Psi}=\frac{1}{(\sum_{j,k=1}^{2}c_{j}^{\star}c_{k}\braket{\psi_{j}}{\psi_{k}})^{1/2}}\sum_{i=1}^{2}c_{i}\ket{\psi_{i}}.
\end{equation}

After taking the expectation value of both sides of eq. (3.28), it can be expressed as a matrix equation
\begin{equation}
\begin{pmatrix}
c_{1}^{\star} & c_{2}^{\star}
\end{pmatrix}
\begin{pmatrix}
p(1|1) & -\braket{\psi_{1}}{\psi_{2}} \\
-\braket{\psi_{2}}{\psi_{1}} & p(2|2)
\end{pmatrix}
\begin{pmatrix}
c_{1} \\
c_{2}
\end{pmatrix} \leq 1
\end{equation}
where the off diagonal terms arise due to the non-trivial normalization of the state $\ket{\Psi}$. The statement that $E_{1}+E_{2} \leq \mathds{1}$ can be rewritten as $\mathds{1}-E_{1}-E_{2} \geq 0$. Using Sylvester's rule, for $\mathds{1}-E_{1}-E_{2} \geq 0$, the element on the first row and first column and the determinant of the matrix must be greater than or equal to zero. Since $(\mathds{1}-M_{1}-M_{2})_{11})$ is a probability, the determinant of the matrix being greater than or equal to zero will suffice.
\begin{equation}
(1-p(1|1))(1-p(2|2)) \geq \abs{\braket{\psi_{1}}{\psi_{2}}}^{2}.
\end{equation}
Note that $p_{i}^{F} = 1-p(i|i)$ due to the law of probability and using it, the constraint on individual failure probabilities can be found as
\begin{equation}
p_{1}^{F}p_{2}^{F} \geq \abs{\braket{\psi_{1}}{\psi_{2}}}^{2}.
\end{equation}
To find the minimum attainable probability for failure, $p_{F}=p_{1}p_{1}^{F}+p_{2}p_{2}^{F}$ must be minimized with respect to the constraint in eq. (3.32). The minimum of $p_{F}$ happens when the product $p_{1}^{F}p_{2}^{F}$ is at its lowest possible value $p_{1}^{F}p_{2}^{F}=\abs{\braket{\psi_{1}}{\psi_{2}}}^{2}$. This allows us to express the individual failure probabilities in terms of the other one
\begin{equation}
p_{2}^{F}=\frac{\abs{\braket{\psi_{1}}{\psi_{2}}}^{2}}{p_{1}^{F}}.
\end{equation}
Inserting this expression into the total failure probability, $p_{1}^{F}$ can be treated as an independent variable and $p_{F}$ can be easily extremized. Doing the same procedure for $p_{1}^{F}$ and $p_{2}^{F}$ gives
\begin{equation}
\begin{split}
p_{1}^{F} &=\sqrt{\frac{p_{2}}{p_{1}}}\abs{\braket{\psi_{1}}{\psi_{2}}}^{2}, \\ p_{2}^{F} &= \sqrt{\frac{p_{1}}{p_{2}}}\abs{\braket{\psi_{1}}{\psi_{2}}}^{2}.
\end{split}
\end{equation}
These values, when inserted into the general expression for the total failure probability $p_{F}=p_{1}p_{1}^{F}+p_{2}p_{2}^{F}$ will yield
\begin{equation}
p_{F}=2\sqrt{p_{1}p_{2}}\abs{\braket{\psi_{1}}{\psi_{2}}}.
\end{equation}
This bound is also known as the IDP bound after Ivanovic, Dieks and Peres. Note that this bound is only applicable if a POVM exists. The existence of the POVM depends on the eigenvalues of the failure operator  which are the individual failure probabilities and they both must satisfy $p_{1}^{F} \leq 1$ and $p_{2}^{F} \leq 1$. This condition can be transformed into an inequality depending on the initial properties of the ensemble like the probabilities and the overlap. The existence range of the POVM is the interval
\begin{equation}
\frac{\abs{\braket{\psi_{1}}{\psi_{2}}}^{2}}{1+\abs{\braket{\psi_{1}}{\psi_{2}}}^{2}} \leq p_{1} \leq \frac{1}{1+\abs{\braket{\psi_{1}}{\psi_{2}}}^{2}}.
\end{equation}
This fact can be seen by examining the expressions for the optimal values of $p_{1}^{F}$ and $p_{2}^{F}$ after making the substitution $p_{2}=1-p_{1}$. If the inequality in eq. (3.36) is not satisfied, a POVM with elements $E_{1}$, $E_{2}$ and $E_{0}$ cannot be constructed and a projective measurement which will be explained below must be used instead.
Using the conjugate states $\ket{\psi_{1}^{\perp}}$ and $\ket{\psi_{2}^{\perp}}$, a measurement scheme can be constructed as follows. Suppose a measurement is performed using two projectors $P_{1}=\dyad{\psi_{1}}$ and $\bar{P}_{1}=\dyad{\psi_{1}^{\perp}}$, and the operators corresponding to $\ket{\psi_{2}}$ and $\ket{\psi_{2}^{\perp}}$. When the measurement $\{P_{1}, \bar{P}_{1}\}$ is performed, if the outcome $\bar{1}$ is obtained then the conclusion that the state was originally $\ket{\psi_{2}}$ can be made without error, but if the outcome $1$ is inconclusive where the same idea can also be applied for the state $\ket{\psi_{2}}$. The probability of the inconclusive result for the measurement $\{P_{1}, \bar{P}_{1}\}$ is simply
\begin{equation}
p_{F1}=p_{1}+p_{2}\abs{\braket{\psi_{1}}{\psi_{2}}}^{2}.
\end{equation}
In the above equation, the $p_{2}$ term corresponds to $\expval{P_{1}}{\psi_{1}}$, which has a probability of $p_{1}$ and the second term corresponds to the $\expval{P_{1}}{\psi_{2}}$ which has a probability of $p_{2}$. The same argument can be applied to the probability of an inconclusive result for the measurement $\{P_{2}, \bar{P}_{2}\}$
\begin{equation}
p_{F2}=p_{2}+p_{1}\abs{\braket{\psi_{1}}{\psi_{2}}}^{2}.
\end{equation}
In total, there are three measurement options for UQSD, the POVM with elements $\{E_{1},E_{2},E_{0} \}$, and the projective measurements $\{P_{1}, \bar{P}_{1}\}$ and $\{P_{2}, \bar{P}_{2}\}$. By checking the values of the failure probabilities for different values of $p_{1}$ while keeping the overlap between the states the same, it can be seen that when $p_{1}$ satisfies the inequality in eq. (3.36), POVM is optimal, when $p_{1}$ is smaller than the lower bound in eq. (3.36) the first projective measurement is optimal and when $p_{1}$ exceeds the upper bound, the second projective measurement is optimal.

The UQSD scheme can be generalized to a procedure what is known as QSS (Quantum State Separation). QSS is a procedure which aims to decrease the square overlaps of a given set of states, in other words $\abs{\braket{\psi^{\prime}_{i}}{\psi^{\prime}_{j}}}^{2} \leq \abs{\braket{\psi_{i}}{\psi_{j}}}^{2}$ is the procedure is successful. The construction is straightforward; a QSS procedure can be thought of as a state transformation problem, transforming the initial set of states $\{\ket{\psi_{i}}\}$ with the overlap matrix $A_{ij}=\braket{\psi_{i}}{\psi_{j}}$ into the set $\{\ket{\psi_{i}^{\prime}}\}$ with the overlap matrix $A_{ij}^{\prime}=\braket{\psi_{i}^{\prime}}{\psi_{j}^{\prime}}$. If $A_{ij} \neq A_{ij}^{\prime}$, this transformation can only be realized with a finite probability of failure if $\abs{\braket{\psi^{\prime}_{i}}{\psi^{\prime}_{j}}}^{2} \leq \abs{\braket{\psi_{i}}{\psi_{j}}}^{2}$ corresponding to the probability of inconclusive result for QSS. If the inner products are to be made smaller, it can be performed without any probability of failure. In this context, UQSD is just a QSS procedure where the final overlap matrix $A^{\prime}_{ij}=\delta_{ij}$. If the transformation is successful, a set of states with an overlap matrix of $\delta_{ij}$ means that the states are mutually orthogonal and discrimination can be realized trivially, via simple orthogonal projective measurements. 
The measurement that is going to perform the state separation procedure is defined as follows, $\mathcal{M}=\{M_{S\lambda},M_{F\lambda}\}$ where S denotes success and F denotes failure and the operators act in the following way
\begin{equation}
\begin{split}
M_{S\lambda}\ket{\psi_{i}} &= c_{\lambda i}\ket{\psi_{i}^{\prime}}, \\ M_{F\lambda}\ket{\psi_{i}} &= f_{\lambda i}\ket{\phi_{\lambda i}}
\end{split}
\end{equation}
for some $\ket{\phi_{\lambda i}}$ and the conditional probabilities are defined as
\begin{equation}
\begin{split}
p(\lambda |i) &= \expval{M_{S\lambda}^{\dagger}M_{S\lambda}}{\psi_{i}}, \\ p^{F}(\lambda |i)&=\expval{M_{F\lambda}^{\dagger}M_{F\lambda}}{\psi_{i}}.
\end{split}
\end{equation}
The measurement $\mathcal{M}$ must obey the completeness relation
\begin{equation}
\sum_{\lambda}M_{S\lambda}^{\dagger}M_{S\lambda}+\sum_{\lambda}M_{F\lambda}^{\dagger}M_{F\lambda} = \mathds{1}
\end{equation}
and the matrix elements of this matrix equation can be found easily
\begin{equation}
\begin{split}
\sum_{\lambda}\bra{\psi_{i}}M_{S\lambda}^{\dagger}M_{S\lambda}\ket{\psi_{j}}+\sum_{\lambda}\bra{\psi_{i}}M_{F\lambda}^{\dagger}M_{F\lambda}\ket{\psi_{j}} &= \braket{\psi_{i}}{\psi_{j}}, \\ 
\sum_{\lambda}c_{\lambda i}^{\star}c_{\lambda i}\braket{\psi_{i}^{\prime}}{\psi_{j}^{\prime}} + \sum_{\lambda}f_{\lambda i}^{\star}f_{\lambda i}\braket{\phi_{\lambda i}}{\phi_{\lambda j}} &= \braket{\psi_{i}}{\psi_{j}}, \\ \sum_{\lambda}c_{\lambda i}^{\star}c_{\lambda i}A_{ij}^{\prime} + \sum_{\lambda}f_{\lambda i}^{\star}f_{\lambda i}\braket{\phi_{\lambda i}}{\phi_{\lambda j}} &= A_{ij}.
\end{split}
\end{equation}
The following definitions are made to express the above equation in a more compact form
\begin{equation}
K_{ij}=\sum_{\lambda}c_{\lambda i}^{\star}c_{\lambda i}, \, \, F_{ij}=\sum_{\lambda}f_{\lambda i}^{\star}f_{\lambda i}\braket{\phi_{\lambda i}}{\phi_{\lambda j}}.
\end{equation}
Using these definitions, the last line of eq. (3.42) can be expressed as
\begin{equation}
A_{ij} = K_{ij}A_{ij}^{\prime}+F_{ij}
\end{equation}
or in matrix form
\begin{equation}
A=K \circ A^{\prime} + F
\end{equation}
where $\circ $ denotes the entry-wise product also known as the Hadamard product. Since the matrices $K$ and $F$ are related to success and failure probabilities, there is a constraint that both $K$ and $F$ must be positive semi-definite. To define the operators themselves, the conjugate set must be defined as
\begin{equation}
\begin{split}
\{\ket{\psi_{i}^{\perp}}\}_{i=1}^{N}, \, \, \text{where} \, \, \ket{\psi_{i}^{\perp}}=\sum_{j}A^{-1}_{ji}\ket{\psi_{i}}, \, \,  \text{and} \, \, \braket{\psi_{k}}{\psi_{i}^{\perp}} &= \sum_{j}A^{-1}_{ji}A_{kj}, \\ &= \delta_{ki}.
\end{split}
\end{equation}
Since $K$ is a hermitian matrix, it can be expressed as follows
\begin{equation}
K=\sum_{\lambda}^{m}a_{\lambda}a^{\dagger}_{\lambda}
\end{equation}
where $a_{\lambda}$ is an unnormalized column matrix and $m$ is greater than or equal to the rank of $K$.
Using these definitions, the operators corresponding to successful state separation can be constructed as follows
\begin{equation}
M_{S\lambda}=\sum_{i}a^{\star}_{\lambda i}\op{\psi_{i}^{\prime}}{\psi_{i}^{\perp}}.
\end{equation}
Checking the effect of $M_{S\lambda}$ on $\ket{\psi_{i}},$ 
\begin{equation}
\begin{split}
M_{S\lambda}\ket{\psi_{i}} &= \sum_{j}a^{\star}_{\lambda j}\ket{\psi_{j}^{\prime}}\braket{\psi_{j}^{\perp}}{\psi_{i}}, \\ &= a^{\star}_{\lambda i}\ket{\psi_{i}^{\prime}}.
\end{split}
\end{equation}
The conditional probability $p(\lambda|i)$ is then $\abs{a_{\lambda i}}^{2}$. Comparing it with the initial definition, it is seen that $c_{\lambda i} = a^{\star}_{\lambda i}$ and the success operators are $M_{S\lambda} = \sum_{i}c_{\lambda i}\ket{\psi_{i}^{\prime}}\bra{\psi_{i}^{\perp}}$. The total probability of successfully transforming the state $\ket{\psi_{i}}$ into $\ket{\psi_{i}^{\prime}}$ is 
\begin{equation}
\begin{split}
p_{s}(i) &= \sum_{\lambda}p(\lambda |i), \\ &=\sum_{\lambda}c^{\star}_{\lambda i}c_{\lambda i}, \\ &= K_{ii}
\end{split}
\end{equation}
meaning that the diagonal elements of the matrix $K$ are the individual probabilities of success for each state. The total probability of success is then simply
\begin{equation}
\begin{split}
p_{S} &= \sum_{i}p_{i}p_{s}(i), \\ p_{S} &= \sum_{i}p_{i}K_{ii}.
\end{split}
\end{equation}
The QSS idea was first put forward by Chefles and Barnett \cite{chefles} for two states and the problem was solved analytically for the special case of two states. Unfortunately, for an arbitrary number of states, this optimization problem is not analytically solvable but it is a nice generalization of the UQSD procedure. 

Up to now, the two main discrimination strategies for monopartite states were covered. A very important problem which has also attracted much attention by the scientific community nowadays is the discrimination of multipartite states, especially under LOCC. In the next sections, perfect discrimination of multipartite, possibly entangled states under LOCC will be covered.

\subsection{Discrimination of Multi-Partite States Under LOCC}
Entanglement plays a central role in the field of quantum information and computation. Entanglement is the cornerstone of many achievements of quantum information theory such as quantum key distribution, quantum teleportation and quantum algorithms. In the context of quantum state discrimination, this translates into the question of the possibility of discriminating between entangled quantum states and the problem has been explored to great depth in the last decade. The problem is not as easy as monopartite state discrimination however and it is filled with negative results. Many authors such as Walgate \cite{walgate}, Virmani \cite{Virmani}, Bandyopadhyay \cite{bandyodisc}, Ghosh \cite{4state} have worked on the subject and found various results about the problem. These results have a common feature, after the discrimination procedure the states are transformed into product states with no entanglement but since entanglement is in the cornerstone of many applications in quantum information theory, in recent years some attention was focused on achieving discrimination with remaining entanglement in the posterior states. Cohen  \cite{cohenent,cohenresource} and Bandyopadhyay \cite{entcost,2qubitcost} have shown that it is indeed possible if the parties agree to use up some entanglement of  previously shared states. Discrimination of entangled states is also closely related with the theory of entanglement transformations and the mathematical theory of majorization. In this section, the results of various authors will be reviewed in depth and the relation of the theory of majorization to the state discrimination problem will be examined.

First, the LOCC procedure should be explained. A LOCC procedure means that the parties sharing a system are only allowed to act on their respective subsystem and share the results of the action by classical communication. It is quite hard to represent a general LOCC procedure generally in a purely mathematical language but a mathematical description can be given as the following. Suppose there are two parties, Alice and Bob and they implement a LOCC scheme. Alice starts the procedure by  a measurement $\mathcal{M}_{A} \otimes \mathds{1}_{B}$ with the possible outcomes labeled by the index $i$. After Alice obtains her outcome, she transmits the result via a classical channel (telephone) to Bob then Bob measures $ \mathds{1}_A \otimes \mathcal{M}_{B}^{i}$ where the $i$ denotes the $i^{th}$ measurement option Bob chooses according to Alice's result. The procedure goes on in the following way until both parties agree to stop. Unfortunately, LOCC procedures are hard to express by pure mathematics and the procedures examined here will usually be explained in a somewhat high-level language.

It is a trivial procedure to discriminate two monopartite orthogonal states as it was shown in the previous sections but it is not that trivial to show that it is possible for multipartite states. To  show that it is indeed possible, it has to be shown that the states in question can always be transformed into a desirable form. Consider the two states
\begin{equation}
\begin{split}
\ket{\psi_{1}} &=\sum_{i}\ket{i}_{A} \ket{\mu_{i}}_{B}, \\
\ket{\psi_{2}}&=\sum_{i}\ket{i}_{A} \ket{\nu_{i}}_{B},
\end{split}
\end{equation}
where the set $\{\ket{i} \}_{i=1}^{N}$ is an orthonormal basis for the Hilbert spaces $\mathcal{H}_{A}$ and $\mathcal{H}_{B}$ but the states $\{\ket{\mu_{i}}\}_{i=1}^{N}$ and $\{\ket{\nu_{i}}\}_{i=1}^{N}$ are not necessarily normalized nor mutually orthogonal. Any bipartite state can be expressed in the form above since it is a general superposition of states where the superposition coefficients have been absorbed in the states of the subsystem $B$. There exists a basis for the Hilbert space $\mathcal{H}_{A}$, $\{\ket{i^{\prime}}\}$ for which the states $\ket{\psi_{1}}$ and $\ket{\psi_{2}}$ take the form 
\begin{equation}
\begin{split}
\ket{\psi_{1}} &=\sum_{i}\ket{i^{\prime}}_{A} \ket{\mu_{i}}_{B}, \\
\ket{\psi_{2}}&=\sum_{i}\ket{i^{\prime}}_{A} \ket{\mu_{i}^{\perp}}_{B},
\end{split}
\end{equation}
where $\braket{{\mu_{i}}}{{\mu_{i}^{\perp}}}=0$, $\{\ket{i^{\prime}}\}_{i=1}^{N}$ is an orthonormal basis for $\mathcal{H}_{A}$, $\{\ket{\mu_{i}}\}_{i=1}^{N}$ and $\{\ket{\mu_{i}^{\perp}}\}_{i=1}^{N}$ are not necessarily orthogonal nor normalized, it is clear that a local projective measurement can perfectly distinguish between the states since after Alice performs a measurement using the projectors $\{\ket{i^{\prime}}_{A}\bra{i^{\prime}}\}_{i=1}^{N}$, Bob can perform a projective measurement using two projectors $P_{1}=\ket{\mu_{i}}_{B}\bra{\mu_{i}}$ and $P_{2}=\ket{\mu_{i}^{\perp}}_{B}\bra{\mu_{i}^{\perp}}$ conditional on the outcome of Alice's measurement $i$. If Bob gets the outcome $1$ than it is certain that the state was $\ket{\psi_{1}}$ and vice-versa. Hence, it is sufficient to show that there always exists a basis $\{\ket{i^\prime}\}$ that enables the states to be expressed as in eq. (3.53).

The states $\ket{\mu_{i}}_{B}$ and $\ket{\nu_{i}}_{B}$ can be expressed as a superposition of the elements of an orthonormal basis for the Hilbert space $\mathcal{H}_{B}$ as
\begin{equation}
\begin{split}
\ket{\mu_{i}}_{B} &= \sum_{j} (\psi_{1})_{ij}\ket{j}_{B}, \\ \ket{\nu_{i}}_{B} &=\sum_{j} (\psi_{2})_{ij} \ket{j}_{B},
\end{split}
\end{equation}
thus, the states $\ket{\psi_{1}}$ and $\ket{\psi_{2}}$ take the form
\begin{equation}
\begin{split}
\ket{\psi_{1}} &= \sum_{i,j}(\psi_{1})_{ij}\ket{i}_{A}\ket{j}_B, \\ \ket{\psi_{2}} &= \sum_{i,j}(\psi_{2})_{ij}\ket{i}_{A}\ket{j}_B.
\end{split}
\end{equation}
The above equation can also be understood in the way that the states in eq. (3.52) are indeed a general form to express two multipartite entangled states. The matrices $\psi_{1}$ and $\psi_{2}$ encode all the necessary information about the states to be discriminated 
\begin{equation}
\begin{split}
\braket{\nu_{i}}{\mu_{j}} &= \sum_{k,l}(\psi_{2})_{ik}^{\star}(\psi_{1})_{jl}\braket{k}{l}, \\ &= \sum_{k}(\psi_{1})_{jk}(\psi_{2})_{ik}^{\star}, \\ &= (\psi_{1}\psi_{2}^{\dagger})_{ji},
\end{split}
\end{equation}
and since $\braket{\psi_{2}}{\psi_{1}} = \sum_{i}\braket{\nu_{i}}{\mu_{i}}$ than if $\braket{\psi_{2}}{\psi_{1}} = 0, \, \, \Tr(\psi_{1} \psi_{2}^{\dagger})=0$. But for discrimination purposes, $\Tr(\psi_{1} \psi_{2}^{\dagger})=0$ condition is not enough. The condition for discrimination is the following
\begin{equation}
\braket{\nu_{i}}{\mu_{i}}=0 \, \, \forall i.
\end{equation}
This means that for some initial traceless matrix $\psi_{1} \psi_{2}^{\dagger}$, if Alice can find a local unitary transformation under which $\psi_{1} \psi_{2}^{\dagger}$ transforms into a matrix with all the diagonal elements 0, discrimination can be achieved. To prove that it is indeed possible, the existence of a unitary transformation that transforms a matrix to an equi-diagonal matrix must be proven. Since $\psi_{1} \psi_{2}^{\dagger}$ is traceless by construction, any transformation that makes it equi-diagonal thus makes it a zero-diagonal matrix. Suppose Alice carries out a unitary transformation (a rotation) on her respective orthonormal basis $\ket{i}_{A}=\sum_{j}(U_{A})_{ij}^{\dagger}\ket{j^{\prime}}_{A}$. Under such a transformation, the state $\ket{\psi_{1}}$ and thus $\ket{\psi_{2}}$ transform as
\begin{equation}
\ket{\psi_{1}}=\sum_{i,j,k}(U_{A})_{ij}^{\dagger}\ket{j^{\prime}}_{A}(\psi_{1})_{ik}\ket{k}_{B}.
\end{equation}
For the sake of generality, the case that Bob also carries out a transformation must also be considered, $\ket{i}_{B}=\sum_{j}(U_{B})_{ij}^{\dagger}\ket{j^{\prime}}_{B}$. Under both transformations, the states transform into
\begin{equation}
\ket{\psi_{1}}=\sum_{i,j,k,l}\ket{j^{\prime}}_{A}\ket{k^{\prime}}_{B}(U_{A})^{\star}_{ji}(\psi_{1})_{ik}(U_{B})_{kl}^{\dagger}.
\end{equation}
From eq. (3.59), it can be concluded that under such unitary transformations, the matrices $\psi_{1}$ and $\psi_{2}$ transform as
\begin{equation}
\psi_{1}^{\prime}=U_{A}^{\star}\psi_{1}U_{B}^{\dagger}, \, \, \psi_{2}^{\prime}=U_{A}^{\star}\psi_{2}U_{B}^{\dagger}.
\end{equation}
The important matrix however is $\psi_{1} \psi_{2}^{\dagger}$. Under such a transformation it becomes
\begin{equation}
\begin{split}
\psi_{1}^{\prime} \psi_{2}^{\prime \dagger} &= U_{A}^{\star}\psi_{1}U_{B}^{\dagger}(U_{A}^{\star}\psi_{2}U_{B}^{\dagger})^{\dagger}, \\ &= U_{A}^{\star}\psi_{1}U_{B}^{\dagger}U_{B}\psi_{2}^{\dagger}U_{A}^{\star \dagger}, \\ &= U_{A}^{\star}\psi_{1}\psi_{2}^{\dagger}U_{A}^{\star \dagger}.
\end{split}
\end{equation}
The above equation tells that such a transformation can be carried out by Alice alone and if $U$ is unitary than $U^{\star}$ is also unitary. Now the proof that such a unitary $U$ can transform $\psi_{1}\psi_{2}^{\dagger}$ into a matrix with equal diagonal elements should be proven. To that end, suppose $\Psi$ is a general $2 \times 2$ matrix and $U$ is a general unitary
\begin{equation}
\Psi = \begin{pmatrix} 
x & y \\
z & t 
\end{pmatrix}, \, \, 
U = \begin{pmatrix}
\cos \theta & \sin \theta e^{i\phi} \\
\sin \theta e^{-i \phi} & -\cos \theta
\end{pmatrix}.
\end{equation}
$U$ is a rotation matrix and it is unitary by construction. The condition that $U\Psi U^{\dagger}$ has equal diagonal elements means that
\begin{equation}
(x-t)\cos 2\theta + (y e^{-i \phi}+z e^{i\phi})\sin 2\theta  = 0.
\end{equation}
As true for any equation with complex variables, this equation can be split into its real and imaginary parts and solved. The real and the imaginary parts of eq. (3.64) are respectively
\begin{equation}
\begin{split}
\Re{x-t}\cos 2\theta + \Re{y+z}\cos \phi \sin 2\theta + \Im{y-z}\sin \phi \sin 2\theta &= 0, \\ \Im{x-t}\cos 2\theta +\Im{y+z}\cos \phi \sin 2\theta + \Re{z-y}\sin \phi \sin 2\theta &= 0.
\end{split}
\end{equation}
Dividing the equations by $\sin 2\theta$ gives
\begin{equation}
\begin{split}
\tan 2\theta &= \frac{\Re{x-t}}{\Re{y+z}\cos \phi - \Im{z-y}\sin \phi}, \\ &= \frac{\Im{x-t}}{\Re{y-z}\sin \phi - \Im{y-z}\cos \phi},
\end{split}
\end{equation}
and combining these two and solving for $\phi$ will give
\begin{equation}
\tan \phi = \frac{\Im{x-t}\Re{z+y}-\Re{x-t}\Im{z+y}}{\Re{x-t}\Re{z-y}+\Im{x-t}\Im{z-y}}.
\end{equation}
Since eq. (3.66) is real, it can always be solved for $\phi$ and thus eq. (3.65) is always solvable for $\theta$. Hence for any $\Psi$, a unitary transformation realized by the unitary matrix $U$ always exists and can be constructed using eqs. (3.65-66). This result however applies to $2 \times 2$ matrices but it can be used for any $2^{n} \times 2^{n}$ matrices since the diagonal elements can be grouped into $2^{n-1}$ pairs and then $2^{n-2}$ quartets can be created and pairs in these quartets can be equalized. Continuing in this manner, any $2^{k} \times 2^{k}$ matrix can be equi-diagonalized by Alice if she applies $k2^{k-1}$ times a unitary $2 \times 2$ transformation. If the matrix $\psi_{1}\psi_{2}^{\dagger}$ is not $2^{k} \times 2^{k}$, an ancilla qubit known to be in the state $\ket{0}_{A^{\prime}}$ can be used to enlarge the dimension of the Hilbert space as follows. The ancilla is introduced in the following manner and  the states $\ket{\psi_{1}}$ and $\ket{\psi_{2}}$ can be expressed as
\begin{equation}
\begin{split}
\ket{\psi_{1}} &=\sum_{i=1}^{n}\ket{i0}_{AA^{\prime}}\ket{\mu_{i}}_{B}+\sum_{i=1}^{n}\ket{i1}_{AA^{\prime}}\ket{\mu_{i+n}}_{B}, \\ \ket{\psi_{2}} &=\sum_{i=1}^{n}\ket{i0}_{AA^{\prime}}\ket{\nu_{i}}_{B}+\sum_{i=1}^{n}\ket{i1}_{AA^{\prime}}\ket{\nu_{i+n}}_{B}.
\end{split}
\end{equation}
Since the state of the ancilla is known to be in $\ket{0}_{A^{\prime}}$, $\ket{\mu_{i}}, \, \, i > n$ and $\ket{\nu_{i}}, \, \, i > n$ have an amplitude of 0. Number theory tells us that $n \leq 2^{k} \leq 2n$ for some $k$, hence Alice can pick a $2^{k} \times 2^{k}$ sub matrix of the expanded $\psi_{1}\psi_{2}^{\dagger}$ and proceed with the equi-diagonalization in the manner described above.

This concludes the proof that any two bipartite orthogonal quantum state can always be perfectly discriminated. The idea can easily be generalized to multipartite states with the party number larger than 2. Using a larger Hilbert space as explained above, the measurement procedure can first be applied by Alice, leaving other parties to continue the discrimination. Than, some other party does the same type of measurement Alice performs and this continues until only 2 parties are left after which the problem reduces to the 2 party version. If they are not orthogonal, the minimum error discrimination strategy can be used. The result is due to Virmani et. al. \cite{Virmani} and an outline of the idea will be presented. If $\braket{\psi_{1}}{\psi_{2}} \neq 0$, than it will not be possible to bring the states into the form in eq. (3.53). The parties however can adopt the following strategy. After Alice performs her projective measurement, Bob will be left with two states conditional on the outcome of Alice. These states, since $\braket{\psi_{1}}{\psi_{2}} \neq 0$, will not be orthogonal but Bob will know the states he possesses. After the measurement of Alice, Bob will be left with two states $\ket{\mu^{\prime}_{i}}$ and $\ket{\nu^{\prime}_{i}}$ conditional on the outcome $i$ of Alice. Now, using eq. (3.56), it can be seen that the elements of the matrix $(\psi_{1}\psi_{2})$ are the inner products, $\braket{\nu_{i}^{\prime}}{\mu_{j}^{\prime}}$. After the equi-diagonalization of $(\psi_{1}\psi_{2})$, all inner products  $\braket{\nu_{i}^{\prime}}{\mu_{i}^{\prime}}$ will be equal. He can then carry out a minimum error discrimination procedure as explained in sec. (2.2) or a UQSD procedure explained in sec. (2.3). This idea shows that two bipartite states can be distinguished even if they are not orthogonal to each other. 

The case for 3 or more states is not as easy to explore as the 2 state case however. Ghosh and his colleagues \cite{4state} have shown that for three bipartite orthogonal pure states, discrimination is only possible for certain intervals and discrimination of 4 mutually orthogonal states is impossible. Their result will be reviewed in detail in the following discussion and its connections with the theory of local entanglement transformations and thus majorization will be explained. 

Consider the set of conjugate states
\begin{equation}
\begin{split}
\ket{\psi_{1}}&=a\ket{00}+b\ket{11},\\
\ket{\psi_{2}}&=b\ket{00}-a\ket{11},\\
\ket{\psi_{3}}&=c\ket{01}+d\ket{10},\\
\ket{\psi_{4}}&=d\ket{00}-c\ket{11},
\end{split}
\end{equation}
where $a,b,c,d$ are chosen to be real without losing any generality. It will be shown in the next discussion that these 4 states cannot be distinguished. To prove it, a few definitions about entanglement monotones must be made. A widely used entanglement monotone is the logarithmic negativity which is defined as 
\begin{equation}
E_{N}(\rho)=\log_{2}\norm{\rho^{T_{A}}}_{1}
\end{equation}
where $\norm{\rho}_{1}$ is the trace norm defined as $\norm{\rho}_{1}=\Tr(\sqrt{\rho^{\dagger}\rho})$ and $\rho^{T_{A}}$ is the partial transpose over the Hilbert space $\mathcal{H}_{A}$ defined in the following way
\begin{equation}
\begin{split}
\rho_{AB}^{T_{A}}&=(\sum_{ijkl}c_{ij}c^{\star}_{kl}\ket{i}_{A}\bra{k}\otimes \ket{j}_{B}\bra{l})^{T_{A}}, \\ &= \sum_{ijkl}c_{ij}c^{\star}_{kl}\ket{k}_{A}\bra{i}\otimes \ket{j}_{B}\bra{l}.
\end{split}
\end{equation}
As defined in chapter 1, entanglement monotones are functions of the Schmidt coefficients of the states such that they don't increase under LOCC. Also, the concept of distillable entanglement is a useful concept and should be defined. Entanglement distillation \cite{distill} is a much used idea in quantum information theory and it is conceptually very simple. Two observers, say Alice and Bob share $n$ entangled states each having less entanglement than a Bell state. Instead of having $n$ copies of a state that has less than 1 ebit of entanglement, using LOCC, these observers can create $m$ Bell states with $m <n$. The procedure can be mathematically outlined in the following way. 

Suppose that Alice and Bob share $n$ copies of a partially entangled state. The state they share can be expressed in the following completely general way
\begin{equation}
\begin{split}
\ket{\Psi}_{AB} &=(\cos \theta \ket{\alpha_{1}\beta_{1}}+\sin \theta\ket{\alpha_{2}\beta_{2}})^{\otimes n}, \\ &= \bigotimes_{i=1}^{n}(\cos \theta \ket{\alpha_{1i}\beta_{1i}}+\sin \theta \ket{\alpha_{2i}\beta_{2i}}), \\ &= \sum_{k=0}^{n}(\cos \theta)^{n-k}(\sin \theta)^{k}[\sum_{\substack{\abs{n=1}=n-k \\ \abs{n=2}=k}}(\bigotimes_{i=1}^{n}\ket{\alpha_{ni},\beta_{ni}})
\end{split}
\end{equation}
where $\otimes n$ denotes the $n^{th}$ power of the state in parenthesis with respect to the Kronecker product and the limits $\abs{n=1}=n-k$ and $\abs{n=2}=k$ means that the index $n$ takes the value 1 for $n-k$ terms and the value 2 for $k$ terms. To obtain a maximally entangled pair, one of the parties will carry out an incomplete projective measurement projecting the states into the subspaces formed by the states having a coefficient of $\sin^{k} \theta$ where $k=0\dots n$ using the projectors
\begin{equation}
P_{k}=\sum_{\substack{\abs{n=1}=n-k \\ \abs{n=2}=k}}(\bigotimes_{i=1}^{n}\dyad{\alpha_{ni}}),
\end{equation}
where these projectors are the ones Alice would use. There are $n+1$ such subspaces and the probability of obtaining the outcome $k$ has a probability of
\begin{equation}
p_{k}= {{n}\choose{k}}(\cos^{2 }\theta)^{n-k}(\sin^{2} \theta )^{k}.
\end{equation}
The combination symbol counts the number of Kronecker product terms with coefficients $(\cos \theta)^{n-k}(\sin \theta)^{k}$ in the expansion of the state $\ket{\Psi}_{AB}$ and it can easily be justified, there are ${{n}\choose{k}}$ ways of choosing $k$ different items out of $n$ items. After learning the result of Alice's measurement, Bob carries out the same projective measurement but due to the entanglement between them, he will always obtain the outcome $k$. Now, note that after Bob's measurement, Alice and Bob will share a maximally entangled state living in a ${{n}\choose{k}}$ dimensional subspace of the original $2^{n}$ dimensional one. Any maximally entangled state can be transformed into a standard form such as a singlet state with LOCC and after such a transformation, Alice and Bob will share a maximally entangled state. The proof will not be taken any further, the interested reader can follow these papers \cite{distill,distill2}.

The property that the logarithmic negativity is the upper bound for distillable entanglement can be used to show that the states $\{\ket{\psi_{i}}\}_{i=1}^{4}$ are not distinguishable. The proof that it is indeed the case is quite long and very non-trivial, so only the reference will be given: the proof can be found in \cite{neg}.

For the sake of notational simplicity, the Bell states will be denoted as $\{\ket{\phi_{i}}\}_{i=1}^{4}$ where the states are ordered from 1 to 4 in the order of appearance in eq. (1.41). Consider the following state that is shared between 4 parties, Alice, Bob, Charlie and Daniel. 
\begin{equation}
\rho_{ABCD}=\frac{1}{4}\sum_{i,j=1}^{4}\ket{\psi_{i}}_{AB}\bra{\psi_{j}} \otimes \ket{\phi_{i}}_{CD}\bra{\phi_{j}}.
\end{equation}
The initial probabilities have been chosen to be equal to make calculations easier and the result more clear. If it were possible to distinguish the states $\ket{\psi_{i}}$ with certainty, Alice and Bob could implement a discrimination scheme after which Charlie and Daniel would share a Bell state which has an entanglement of 1 ebit. This can be interpreted in the context of distillable entanglement to mean that the distillable entanglement between Charlie - Daniel and Alice - Bob, or the AC:BD cut is 1 ebit. Now, use the fact that logarithmic negativity is an upper bound for distillable entanglement. The logarithmic negativity of $\rho_{ABCD}$ in the AC:BD cut is
\begin{equation}
\begin{split}
E_{N}(\rho) &= \log_{2}\norm{\rho^{T_{AC}}}_{1}, \\ &= \log_{2}\Tr((\rho^{T_{AC}})^{\dagger}\rho^{T_{AC}})^{1/2}).
\end{split}
\end{equation}
$\rho$ is given by $\rho = \frac{1}{4}\sum_{i,j=1}^{4}\ket{\psi_{i}}_{AB}\bra{\psi_{j}} \otimes \ket{\phi_{i}}_{CD}\bra{\phi_{j}}$ and when $E_{N}(\rho)$ is calculated it gives
\begin{equation}
E_{N}(\rho) = \log_{2}(a^{2}+c^{2}).
\end{equation}
The terms in the logarith is just squares since it was assumed that $a,b,c,d \in \mathds{R}$. The contradiction is clear. Using the fact that the logarithmic negativity is the upper bound for distillable entanglement, $E_{N}(\rho) \geq 1$, both $a$ and $c$ must be 1, where 1 is the distillable entanglement in the $AC:BD$ cut. If $a=c=1$, the states $\{\ket{\psi_{i}}\}_{i=1}^{4}$ become product states. The conclusion to be drawn here is that 4 entangled, mutually orthogonal states cannot be distinguished. Also, note that when the states $\{\ket{\psi_{i}}\}_{i=1}^{N}$ are chosen as the Bell states, the bound can not be satisfied meaning that the Bell states cannot also be discriminated. If this test is applied to the case where only 3 states of the 4 original ones are to be discriminated, it turns out that discrimination is possible for some range of the parameters $a,b,c,d$. The density operator to be used in the test is constructed as follows
\begin{equation}
\rho_{ABCD}=\frac{1}{4}\sum_{i,j=1}^{3}\ket{\psi_{i}}_{AB}\bra{\psi_{j}} \otimes \ket{\phi_{i}}_{CD}\bra{\phi_{j}},
\end{equation}
and when the logarithmic negativity is calculated for the density operator in eq. (3.77), it gives
\begin{equation}
E_{N}(\rho)=\log_{2}[1+\frac{1}{3}(1+16a^{2}b^{2}-4c^{2}d^{2})^{1/2}+2(1-4a^{2}b^{2}+c^{2}d^{2})^{1/2}].
\end{equation}
For three states to be discriminated, $E_{N}(\rho) \geq 1$ must be satisfied, if else there is a contradiction with the fact that logarithmic negativity is the upper bound for distillable entanglement. The bound is satisfied if 
\begin{equation}
4a^{2}b^{2}-c^{2}d^{2} >\frac{3}{4}.
\end{equation}
A quick note, the constraint in eq. (3.79) is reached when the states $\ket{\psi_{i}}, \, i=1,2,3$ are used or, in a more qualitative description, two parallel and one anti-parallel state is used. Since the Schmidt coefficients are same for $\ket{\psi_{1}}$ and $\ket{\psi_{2}}$, and $\ket{\psi_{3}}$ and for $\ket{\psi_{4}}$, there are only 2 different ways of picking the 3 states to be discriminated among the 4. If two anti parallel and 1 parallel state is chosen, the constraint is found by swapping $a$ with $c$ and $b$ with $d$, giving
\begin{equation}
4c^{2}d^{2}-a^{2}b^{2} > \frac{3}{4}.
\end{equation}
This is a good point to start discussing the theory of majorization and its connections to entanglement transforms and hence the state discrimination problem. In the next section, the theory of majorization will be introduced and its connections to the state discrimination problem will be made clear. 

\subsection{Majorization, Entanglement Transformations and State Discrimination}
The theory of majorization is a widely studied subject in mathematics, particularly in the field of linear algebra. The applications of majorization are widespread, but for quantum information theory, the main attraction of the theory comes from the ability to compare the mixedness of two probability distributions. The connection between quantum information theory and the theory of majorization is explained in great detail in Nielsen's papers \cite{maj1,Nielsen2001}. Majorization is a powerful tool which completely characterizes transformations of pure entangled states and gives important relations about the quantum state of a system before and after measurement and the outcome probabilities. The focus in this section and the rest of this document will be on the first main application of majorization and it will be explained in detail below. 

Majorization defines a partial order between vectors whose elements sum up to the same value. Majorization can be defined in the following way. Let $x$ and $y$ be $d$ dimensional vectors
\begin{equation}
x=\begin{pmatrix}
x_{1} \\
x_{2} \\
\vdots \\
x_{d}
\end{pmatrix}, \, \, y=\begin{pmatrix}
y_{1} \\
y_{2} \\
\vdots \\
y_{d}
\end{pmatrix}.
\end{equation}
Than, $y$ majorizes $x$, or $x \prec y$ if the following inequality holds for every $k = 1,2,\dots,d$ and equality holds for $k=d$
\begin{equation}
\sum_{i=1}^{k}x_{i}^{\downarrow} \leq \sum_{i=1}^{k}y_{i}^{\downarrow},
\end{equation}
where $x_{i}^{\downarrow}$ denotes  the components of the vector $x$ are organized in a descending order and specifically, $x_{i}^{\downarrow}$ is the $i^{th}$ largest component of $x$. This partial order can be extended to matrices with equal traces constructing a vector $\lambda_{A}$ using the eigenvalues of the matrix $A$ and sorting the eigenvalues in a descending order. Note that when two vectors of unequal dimension but equal sum of elements are going to be compared, the lesser dimensional can be padded with zeros until its dimension equals the larger dimensional vector and this idea is going to be used a lot.

The connection of majorization with quantum information theory is made through comparing the eigenvalue vectors of the reduced density operators of various quantum states. There are three important results that realize the connection. These results will be explained in detail below.

The first connection with information theory can be made by using the notion of Schur convex and Schur concave functions. A Schur convex function is a function such that
\begin{equation}
f:\mathds{R}^{d} \rightarrow \mathds{R}^{d}, \, \, f(x) \leq f(y)\, \, \text{iff}\, \, x \prec y, \, \, \forall x,y, \, \in \mathds{R}^{d}.
\end{equation}
If $f$ is Schur convex, than $-f$ is Schur concave. All Schur convex functions are symmetric under the permutations of its arguments. A criteria for Schur convexity can be expressed as follows. Let $f(x_{1},\dots,x_{d})$ be a function on $\mathds{R}^{d}$ with continuous partial derivatives. Then $f$ is Schur convex if
\begin{equation}
(x_{i}-x_{j})(\pdv{f}{x_{i}}-\pdv{f}{x_{j}}) \geq 0, \, \, \forall x_{i},x_{j} \in \mathds{R}^{d}.
\end{equation}
It can be seen from inspection that the function $f(x_{i})=\sum_{i=1}^{d}x_{i}\log_{2}x_{i}$ is Schur convex. Then, the function $H(x_{i})=-\sum_{i=1}^{d}x_{i}\log_{2}x_{i}$ is Schur concave which is also the Shannon entropy function. Remembering the definition of Schur concavity, the connection can be made. If $p(x) \prec p(y)$ where $p(x),p(y)$ are probability distributions, $H(x) \geq H(y)$. In other words, $x$ is more mixed than $y$. Note that, for density operators, the von Neumann entropy reduces to the Shannon entropy with $x_{i}$ becoming the eigenvalues of the density operator. Using majorization, a partial order can be constructed amongst density operators ranking their mixedness. Also remember that the entanglement entropy is defined as the von Neumann entropy of the reduced density operator. Also, using majorization, a partial order among quantum states can be constructed, ranking their entanglement amount. 

Nielsen's theorem and the theorem due to Jonathan and Plenio \cite{JonathanP} produce a fascinating connection between majorization and transformations of pure entangled states under LOCC. Let $\ket{\psi}_{AB}$ and $\ket{\phi}_{AB}$ be two bipartite states and let $\rho_{\psi}=\Tr_{B}\ket{\psi}\bra{\psi}$ and $\rho_{\phi}$ defined accordingly. The theorem states that if $\ket{\psi}$ is can be converted into $\ket{\phi}$ under LOCC, then the following must hold
\begin{equation}
\begin{split}
\ket{\psi} \rightarrow \ket{\phi}\, \, &\text{iff} \, \, \lambda(\psi) \prec \lambda(\phi),\\ \ket{\psi}\rightarrow \{(p_{i},\ket{\phi_{i}}\} \, \, &\text{iff} \, \, \lambda(\psi) \prec \sum_{i}p_{i}\lambda(\phi_{i}),
\end{split}
\end{equation}
where $\lambda(\psi)$ is the vector with the components being the eigenvalues of $\rho_{\psi}$. The second relation is for a probabilistic transformation where the state $\ket{\psi}$ is transformed into $\ket{\phi_{i}}$ with probability $p_{i}.$ There are, however some states for which neither $\lambda(\psi) \prec \lambda(\phi)$ nor $\lambda(\phi) \prec \lambda(\psi)$. These states will be said to have incompatible entanglement. 

$\lambda(\psi)$ can also be seen as a quantifier of entanglement as it does not increase under LOCC. Also, it can be inferred from eq. (3.85) that a maximally entangled state can be transformed to any state with same dimensionality since 
\begin{equation}
\begin{pmatrix}
\frac{1}{d} \\
\frac{1}{d} \\
\vdots \\
\frac{1}{d}
\end{pmatrix} \prec \begin{pmatrix} x_{1} \\ x_{2} \\ \vdots \\ x_{d} \end{pmatrix},
\end{equation}
where $\sum_{i=1}^{d}x_{i}=1$. 
The proof of eq. (3.85) uses several facts which are not going to be proven here:

(i) Any matrix $A$ can be expressed as a polar decomposed form $A=(A^{\dagger}A)^{1/2}U$ where $U$ is some unitary.

(ii) If $\rho^{\prime}=\sum_{i}p_{i}U_{i}\rho U_{i}^{\dagger}$ where $p_{i}$ are probabilities, $\lambda(\rho^{\prime}) \prec \lambda(\rho)$.

(iii) If $x \prec y$, then $x=Dy$ where $D$ is a product of at most $d-1$ $T$ transforms where $d=\dim x$ and a $T$ transforms acts on at most 2 components of a matrix and for those two components it has the form 
\begin{equation}
T=\begin{pmatrix} t & 1-t \\ 1-t & t \end{pmatrix}.
\end{equation}

(iv) $\ket{\psi} \sim \ket{\phi}$ will be used if they have the same Schmidt coefficients. Two states with the same Schmidt coefficients are equivalent under local unitaries.

The LOCC protocol can be thought of the following way; Alice performs a general measurement $\mathcal{M}=\{M_{m}\}$ and Bob performs a quantum operation $\mathcal{E}_m$ conditional on the outcome $m$
\begin{equation}
\ket{\phi}\bra{\phi}=\sum_{m}\mathcal{E}_{m}(M_{m}\ket{\psi}\bra{\psi}M_{m}^{\dagger}).
\end{equation}
A note, Alice and Bob can also realize the transformation using general measurements and arbitrary amount of classical communication or Bob can perform a possibly non unitary operation where Alice and Bob only communicate once. 

Tracing out the subsystem $B$ and noting that the states in question are pure, the relation
\begin{equation}
M_{m}\rho_{\psi}M_{m}^{\dagger}=p_{m}\rho_{\phi}
\end{equation}
can be obtained. Using fact (i) and $\rho \geq 0$ for any density operator
\begin{equation}
\begin{split}
M_{m} \sqrt{\rho_{\psi}} &= (M_{m}\rho_{m}M_{m}^{\dagger})^{1/2}U_{m}, \\ &= \sqrt{p_{m}\rho_{\phi}}U_{m}.
\end{split}
\end{equation}
Since the measurement $\mathcal{M}$ is complete $\rho_{\psi}$ can be expressed as 
\begin{equation}
\rho_{\psi}=\sum_{m}\sqrt{\rho_{\psi}}M_{m}M_{m}^{\dagger}\sqrt{\rho_{\psi}}.
\end{equation}
Now, eq. (3.90) can be substituted in eq. (3.91) to get
\begin{equation}
\rho_{\psi}=\sum_{m}p_{m}U_{m}\rho_{\phi}U_{m}^{\dagger}
\end{equation}
and using fact (ii), $\lambda(\psi) \prec \lambda(\phi)$. This shows that if $\ket{\psi} \rightarrow \ket{\phi}$ then $\lambda(\psi) \prec \lambda(\phi)$. Now the converse statement, if  $\lambda(\psi) \prec \lambda(\phi)$, then $\ket{\psi} \rightarrow \ket{\phi}$ must be proven. 

Without losing any generality, the quantum states $\ket{\psi}$ and $\ket{\phi}$ can be written in their Schmidt decomposed form using the computational basis
\begin{equation}
\begin{split}
\ket{\psi} &\sim \ket{\psi^{\prime}} = \sqrt{p_{1}}\ket{00} + \sqrt{p_{2}}\ket{11}, \\ \ket{\phi} &\sim \ket{\phi^{\prime}} = \sqrt{q_1}\ket{00} + \sqrt{q_{2}}\ket{11},
\end{split}
\end{equation}
where it is assumed that $p_{1} \geq p_{2}$ and $q_{1} \geq q_{2}$, and to satisfy the majorization relation $p_{2} \geq q_{2}$ and $p_{1} \leq q_{1}$. 

Alice and Bob will first try to transform the state $\ket{\psi^{\prime}}$ to
\begin{equation}
\ket{\psi^{\prime \prime}} = \frac{1}{\sqrt{2}}(\ket{00}+\ket{1}(\cos \alpha \ket{0} + \sin \alpha \ket{1})).
\end{equation}
For this to be possible, $\alpha $ must satisfy $p_{1}=\frac{1}{2}(1+\cos \alpha)$ so that the states have the same Schmidt coefficients. Then, Alice performs a general measurement defined through the operators
\begin{equation}
M_{1}=\begin{pmatrix} \cos \beta & 0 \\ 0 & \sin \beta \end{pmatrix}, \, \, M_{2}=\begin{pmatrix} \sin \beta & 0 \\ 0 & \cos \beta \end{pmatrix}.
\end{equation}
It can be checked that $M_{1}^{\dagger}M_{1} + M_{2}^{\dagger}M_{2} = \mathds{1}$. The post measurement states corresponding to $M_{1}$ and $M_{2}$ are
\begin{equation}
\begin{split}
\ket{\psi_{1}^{\prime \prime \prime}} &= \cos \beta \ket{00} + \sin \beta \ket{1}(\cos \alpha \ket{0} + \sin \alpha \ket{1}), \\ \ket{\psi_{2}^{\prime \prime \prime}} &= \sin \beta \ket{00} + \cos \beta \ket{1}(\cos \alpha \ket{0} + \sin \alpha \ket{1}).
\end{split}
\end{equation}
Note that the operators $M_{1}$ and $M_{2}$ are constructed in such a way that the Schmidt coefficients of $\ket{\psi_{1}^{\prime \prime \prime}}$ and $\ket{\psi_{2}^{\prime \prime \prime}}$ are the same. Using fact (iv), it is possible to ensure that the post measurement state is $\ket{\psi_{1}^{\prime \prime \prime}}$ after some local unitary transformations and classical communication. The Schmidt coefficients of the state $\ket{\psi_{1}^{\prime \prime \prime}}$ can be calculated easily
\begin{equation}
\lambda_{\pm}=\frac{1 \pm (1-\sin^{2} 2\beta \sin^{2} \alpha)^{1/2}}{2}.
\end{equation}
If there exists a solution such that $\lambda_{+} = q_{1}$, it means that the state $\ket{\psi_{1}^{\prime \prime \prime}}$ is equivalent to $\ket{\phi^{\prime}}$ under unitary transformations. This equation can be solved easily and gives
\begin{equation}
\beta = \frac{1}{2}\arcsin[\frac{2(q_{1}-q_{1}^{2})}{\sin \alpha}].
\end{equation}
For systems of dimensionality greater than 2, the procedure is to apply $M_{1}^{i}$ and $M_{2}^{i}$ where $i$ refers to a block of two components and acting them in succession to the states
\begin{equation}
\ket{\psi} \sim \ket{\psi^{\prime}} = \cos \gamma (\sqrt{p_{1}}\ket{00} + \sqrt{p_{2}}\ket{11}) + \sin \gamma \ket{\psi^{\perp}}
\end{equation}
and $\ket{\phi^{\prime}}$ is expressed in the same fashion.

This completes the proof of the claim in eq. (3.85). For the proof of the second relation
\begin{equation}
\lambda(\psi) \prec \sum_{i}p_{i}\lambda(\phi_{i}),
\end{equation}
some theorems in the theory of majorization will be used without proof \cite{majrev}. 

(i)Ky Fan's maximum principle: $\sum_{j=1}^{k}\lambda_{j}(A) = \max_{P}\Tr(AP)$ where $P$ are $k$-dimensional projectors.

(ii)$\lambda(A+B)\prec \lambda(A)+\lambda(B)$ which is a consequence of the Ky Fan's maximum principle.

An immediate result of (ii) is for $\rho=\sum_{i}p_{i}\rho_{i},$ 
\begin{equation}
\lambda(\rho) \prec \sum_{i}p_{i}\lambda(\rho_{i}).
\end{equation}

To prove the relation in eq. (3.100), first, it will be proved that if a measurement $M_{i}$ transforms $\ket{\psi}$ into $\ket{\phi_{i}}$ with probability $p_{i}$ then $\lambda(\psi) \prec \sum_{i}p_{i}\lambda(\phi_{i})$ must hold. 

Suppose that the measurement is performed locally on a subsystem of a pure state $\ket{\psi}$ where $\rho_{A}=\Tr_{B}\dyad{\psi}$. The posterior states are: \begin{equation}
\begin{split}
\ket{\psi_{i}}&=\frac{(M_{i}\otimes \mathds{1}_{B})\ket{\psi}}{\sqrt{p_{i}}},\\ \rho_{B,i}&=\Tr_{A}\dyad{\psi_{i}}.
\end{split}
\end{equation}
No communication theorem prevents the faster than light propagation of information or in other words, it prevents Bob from learning whether Alice has performed a measurement or nor or the outcome of Alice's measurement without Alice telling him. The statement can be expressed as follows
\begin{equation}
\begin{split}
\rho_{B}=\Tr_{A}\dyad{\psi} &= \Tr_{A}[(M_{i}\otimes \mathds{1}_{B})\dyad{\psi}(M_{i}^{\dagger}\otimes \mathds{1}_{B})],\\&= \sum_{i}p_{i}\rho_{B,i}.
\end{split}
\end{equation}
Note that the statement above is equivalent to the one in eq. (3.101) since all reduced density operators of a pure state have identical spectrum.

Now suppose that $\rho$ and $\rho_{i}$ are density operators and $p_{i}$ are probabilities such that eq. (3.101) holds. It will be shown that there exists a transformation 
\begin{equation}
\sum_{i,j}M^{\dagger}_{ij}M_{ij} = \mathds{1}, \, \, M_{ij}\rho M_{ij}^{\dagger} = p_{ij}\rho_{i} \, \, \text{and}\, \, \sum_{j}p_{ij}=p_{i}.
\end{equation}

Fact (iii) can be re expressed in the following way. $\lambda(\rho) = \sum_{i,j}p_{i}q_{j}P_{j}\lambda(\rho_{i})$ if $\lambda(\rho) \prec \sum_{i}p_{i}\lambda(\rho_{i})$ where $P_{j}$ are permutation matrices. It is easy to justify since $T$ transformations are convex combinations of permutation matrices. 

The operators $M_{ij}$ are defined as follows
\begin{equation}
M_{ij}\sqrt{\rho}=\sqrt{p_{i}q_{j}}\sqrt{\rho_{i}}P_{j}^{\dagger}
\end{equation}
after assuming that $\rho$ and $\rho_{i}$ are diagonalizable in the same basis and their eigenvalues are sorted decreasingly. This can be achieved by local unitary transformations before the measurement $M_{ij}$.
\begin{equation}
\sum_{i,j}\sqrt{\rho}(M_{ij}^{\dagger}M_{ij})\sqrt{\rho}=\sum_{ij}p_{i}q_{j}P_{j}\rho_{i}P_{j}^{\dagger}.
\end{equation}
Since $\rho$ and $\rho_{i}$ are diagonal and the diagonal elements are in decreasing order, the above equation is equivalent to $\lambda(\rho)=\sum_{i,j}p_{i}q_{j}P_{j}\lambda(\rho_{i})$ meaning that $M_{ij}$ satisfy the completeness relation. 

It also follows that
\begin{equation}
M_{ij}\rho M_{ij}^{\dagger} = p_{i}q_{j}\rho_{i} 
\end{equation}
where $p_{ij}=p_{i}q_{j}$ and $\sum_{j}p_{i}q_{j}=p_{i}$. Combining with the proof of the converse relation, the second theorem is proved.

Using eq. (3.100), the feasibility of state discrimination problems can be investigated by the theory of majorization. Majorization relations can be found for state discrimination problems and the relations will help in understanding whether discrimination of a given set of states is possible or not. However, since majorization covers all kinds of LOCC procedures and does not limit itself to one-way communication, constructing a protocol and the measurement operators that do the job is quite hard in most cases. 

The result of Ghosh et. al. can be investigated using majorization using eq. (3.100). The set of states to be discriminated are the ones in eq. (3.68), and $\ket{\phi_{i}}$ are the Bell states. Construct the multipartite state
\begin{equation}
\ket{\Psi}=\frac{1}{2}\sum_{i=1}^{4}\ket{\psi_{i}}_{AB}\ket{\phi_{i}}_{CD}.
\end{equation}  
If discrimination is successful, The parties $C,D$ will share a maximally entangled state. This procedure can be thought of as an entanglement transformation under LOCC with the majorization relation
\begin{equation}
\lambda(\Tr_{BD}\dyad{\Psi}) \prec \lambda(\frac{1}{2}\mathds{1}).
\end{equation}
This is actually the majorization relation for an entanglement transformation that takes the the state $\ket{\Psi}$ and converts it into a Bell state $\ket{\phi_{i}}$ with probability $p_{i}=\frac{1}{4}$ and $i$ is identified. Being able to identify the index $i$ means that the states $\ket{\psi_{i}}$ are distinguished from each other and the parties $C$ and $D$ now know which Bell state they share. The equation above is just eq. (3.100) with relevant information inserted. The partial trace can be calculated and gives
\begin{equation}
\Tr_{BD}\dyad{\Psi}= \frac{1}{8} \begin{psmallmatrix} (a+b)^{2}+(c+d)^{2} & 0 & 0 & 2(a+b)(c+d) \\ 0 & (a-b)^{2}+(c-d)^{2} & 2(a-b)(c-d) & 0 \\ 0 & 2(a-b)(c-d) & (a-b)^{2}+(c-d)^{2} & 0 \\  2(a+b)(c+d) & 0 & 0 & (a+b)^{2}+(c+d)^{2} \end{psmallmatrix}
\end{equation}
with the eigenvalues 
\begin{equation}
\begin{split}
\lambda_{1} &= \frac{1}{8}(a+b+c+d)^{2}, \\ \lambda_{2} &= \frac{1}{8}(a-b+c-d)^{2}, \\ \lambda_{3} &= \frac{1}{8}(a-b-c+d)^{2}, \\ \lambda_{4} &= \frac{1}{8}(a+b-c-d)^{2}.
\end{split}
\end{equation}
Without losing any generality, it can be assumed that $a \geq b$ and $c \geq d$ and $a,b,c,d \geq 0$. The largest among the eigenvalues is surely $\lambda_{1}$ and for the majorization relation to hold, the following must be satisfied.
\begin{equation}
\frac{1}{2} \geq \frac{1}{8}(a+b+c+d)^{2}.
\end{equation}
The function $a+\sqrt{1-a^{2}}-1$ has only one root in the interval $a \in [0,1]$, it is $a=0$ and that point is also the local minimum hence the function $a+\sqrt{1-a^{2}}-1$ is always greater than or equal to zero with the equality satisfied at the endpoints hence $a+\sqrt{1-a^{2}} \geq 1$. This means that the inequality in eq. (3.112) can only be satisfied for $a=c=0$ or $a=c=1$ and that means that the states $\ket{\psi_{i}}$ are unentangled. The result that 4 mutually orthogonal entangled pure states cannot be distinguished with LOCC. The same idea can be applied for the three state case where 
\begin{equation}
\ket{\Psi}=\frac{1}{\sqrt{3}}\sum_{i=1}^{3}\ket{\psi_{i}}_{AB}\ket{\phi_{i}}_{CD}
\end{equation}
and the majorization relation is as in eq. (3.109). The reduced density matrix for the state $\ket{\Psi}$ becomes
\begin{equation}
\Tr_{BD}\dyad{\Psi} = \frac{1}{6} \begin{psmallmatrix} (a+b)^{2}+c^{2} & 0 & 0 & (a+b)(c+d) \\ 0 & (a-b)^{2}+c^{2} & (a-b)(d-c) & 0 \\ 0 & (a-b)(d-c) & (a-b)^{2}-d^{2} & 0 \\  (a+b)(c+d) & 0 & 0 & (a+b)^{2}+d^{2} \end{psmallmatrix}.
\end{equation}
The eigenvalue expressions are not simple but numerical calculations show that discrimination is possible for a range of values for $a$ and $c$.

Majorization can also be used to investigate a different problem. Suppose Alice and Bob share the states $ \ket{\psi_{i}}$ and they also share an entangled pair
\begin{equation}
\ket{\phi} = \alpha \ket{00} + \beta  \ket{11}
\end{equation}
and they can use the entanglement of $\ket{\phi}$ to achieve discrimination of the states $\ket{\psi_{i}}$. If the parties are able to distinguish between the individual $\ket{\psi_{i}}$, then they will be able to implement a probabilistic entanglement transformation that takes the state $\ket{\phi}\otimes \ket{\Psi}$ to a Bell state $\ket{\phi_{i}}$ with equal probability for each $i$ and identifying the index $i$. Using this idea, a majorization relation for that transformation constructed using eq. (3.100). 
\begin{equation}
\lambda(\Tr_{B}\dyad{\phi}) \otimes \lambda(\Tr_{BD}\dyad{\Psi}) \prec \lambda(\frac{1}{2}\mathds{1}),
\end{equation}
where $\ket{\Psi}$ is defined as before and $\alpha \geq \beta $ is assumed. Using the eigenvalues of $\Tr_{BD}\ket{\Psi}\bra{\Psi}$, the first two terms of the vector in the left hand side of eq. (3.109) is
\begin{equation}
\lambda(\Tr_{B}\dyad{\phi}) \otimes \lambda(\Tr_{BD}\dyad{\Psi})=\frac{1}{8} \begin{pmatrix} \alpha^{2} (a+b+c+d)^{2} \\ \alpha^{2}(a-b+c-d)^{2} \\ \vdots \end{pmatrix}.
\end{equation}
Note that the sum of the first two terms are always smaller than or equal to one hence for the majorization relation to hold
\begin{equation}
\begin{split}
\frac{1}{8}\alpha^{2} (a+b+c+d)^{2} &\leq \frac{1}{2}, \\ \alpha^{2} &\leq \frac{4}{(a+b+c+d)^{2}}.
\end{split}
\end{equation}
The limits of this equation can be checked and gives expected results. If the states $\ket{\psi_{i}}$ are maximally entangled, $\ket{\phi}$ is maximally entangled too and if $\ket{\psi_{i}}$ are product states, $\ket{\phi}$ is not needed to succeed. The amount of entanglement that $\ket{\phi}$ should have for successful discrimination can be calculated using the entanglement entropy formula and for $\ket{\psi_{i}}$ having equal prior probabilities, numerical results show that the entanglement of $\ket{\phi}$ is always larger than the average entanglement of the states. An analytical solution however for the entanglement of $\ket{\phi}$ in terms of the average entanglement is not possible.  Also, a one-way protocol couldn't be found that realizes these results and it is believed that a one-way protocol in which both parties carry out generalized measurements is impossible for the above relation and it is nearly impossible to express all the possible LOCC procedures in a mathematical way. However, this result is still valid in the sense that it shows that entanglement can be used as a resource to perform otherwise impossible results.

This concludes the discussion of quantum state discrimination. In this chapter, various strategies and different cases of the problem has been examined and it has been shown that entanglement can be used to help discriminate states. In the next chapter, the possibility of preserving the entanglement of the states during discrimination will be discussed.

\section{Discrimination with Remaining Entanglement}
In many applications of quantum information theory, the entanglement of the quantum state is of crucial importance and it has many uses. It is therefore only natural to try to preserve the entanglement under various procedures. In the context of quantum state discrimination, this means answering the following question; is it possible to achieve discrimination of quantum states while preserving the entanglement? The answer is, yes if the parties are willing to use the entanglement of a preshared state. This is proven in a number of papers in literature, examples can be found in Cohen's works like \cite{cohenent} and Cohen also proves that for some specific sets of product states called as unextendible product bases,  entanglement is also necessary to succeed \cite{cohenresource}.

The procedure of achieving discrimination with remaining entanglement also has a very intimate connection with the ability to realize non-local measurements. The connection is quite easy to grasp, imagine a set of multipartite entangled states $\ket{\psi_{i}}$, if the parties can come together and implement a projective measurement using the projectors $P_{i}=\dyad{\psi_{i}}$ and then separate, the states will remain entangled. If they cannot physically come together, one of the parties can teleport his part to the other party and they can realize a projective measurement projecting onto the states. This protocol will cost 1 ebit of entanglement, but a protocol using up less entanglement are possible \cite{entcost}. 

For a procedure achieving discrimination with entanglement preservation, the upper bound on the cost is 2 ebits since the most inefficient way to achieve it is to perform 2 quantum teleportations. Alice teleports her part of the states to Bob using 1 ebit of entanglement, Bob performs a projective measurement and teleports the respective part of the state to Alice, recovering the original state.
 
In this chapter, the lower bounds of entanglement for various cases will be calculated using techniques of majorization and the results will be discussed. In some cases, instead of the direct entanglement amounts, the bounds on the Schmidt coefficients of the preshared entangled state will be given and the values corresponding to the maximally entangled and product states will be discussed. As in the previous chapter, the protocols will not be constructed, only the inequalities that a protocol which can realize the discrimination will be given. 

In the context of this work, preservation of entanglement is understood in the sense that the amount of entanglement the state has is the same after discrimination which means that the posterior states have the same Schmidt coefficients as the initial states. Two states with the same Schmidt coefficients are equivalent upto local unitary transformations or basis changes meaning that the posterior states can be rotated into the initial states without losing any entanglement. 

The way to calculate the bounds on Schmidt coefficients using majorization is very similar to the calculation performed at the last section of chapter 2. The main difference lies in the posterior states. If the entanglement remaining in the states is of no concern, the calculation can proceed as in before otherwise, the relation has to be modified. 

This idea can be illustrated in a simple example. Suppose Alice and Bob want to discriminate between the states
\begin{equation}
\begin{split}
\ket{\psi_{1}}&=a\ket{00}+b\ket{11},\\
\ket{\psi_{2}}&=c\ket{01}+d\ket{10},
\end{split}
\end{equation}
where $a,b,c,d$ are assumed to be real and $a \geq b$ and $c \geq d$. Alice and Bob share an entangled state
\begin{equation}
\ket{\phi}=\alpha \ket{00} + \beta \ket{11}.
\end{equation}
A quick note, the only important parameters for the state $\ket{\phi}$ are the Schmidt coefficients. The choice of basis is not important since a simple rotation will transform between two different orthonormal bases. 

To construct a majorization relation that completely characterizes such a procedure, it must be seen as an entanglement transformation process. If the parties can discriminate between the states, the following probabilistic entanglement transformation is implementable.
\begin{equation}
\ket{\phi}\otimes (\sqrt{p_{1}}\ket{\psi_{1}} + \sqrt{p_{2}}\ket{\psi_{2}}) \rightarrow \ket{\psi_{i}}, \, \, \text{with probability}\, \, p_{i}.
\end{equation}
The majorization relation then can be constructed as follows
\begin{equation}
\lambda(\phi) \otimes \lambda(\sqrt{p_{1}}\ket{\psi_{1}}+\sqrt{p_{2}}\ket{\psi_{2}}) \prec p_{1}\lambda(\psi_{1})+p_{2}\lambda(p_{2}).
\end{equation}
To find the lowest possible bound on the Schmidt coefficients of $\ket{\phi}$ The state 
\begin{equation}
\ket{\Psi}=\sqrt{p_{1}}\ket{\psi_{1}}+\sqrt{p_{2}}\ket{\psi_{2}}
\end{equation}
should be made a product state by a proper choice of the coefficients $\sqrt{p_{1}},\sqrt{p_{2}}$. The reason is as follows; in the majorization relation, the coefficients of $\ket{\phi}$ will be multiplied by the coefficients of $\ket{\Psi}$, getting them closer to 0 and closer to each other which means higher entanglement. Therefore, to find the lowest possible bound, the state $\ket{\Psi}$ must be made product. In the first chapter, it was proven that if the reduced density matrix of a quantum state is rank 1, than the corresponding state will be a product state. Since the Hilbert spaces $\mathcal{H}_{A}$ and $\mathcal{H}_{B}$ are two dimensional, the reduced density matrix of $\ket{\Psi}$ will be $2 \times 2$. If the determinant of a $2 \times 2$ matrix vanishes, it means that the matrix in question is a rank 1 matrix and for reduced density operators, it means that the state is a product state. 

The reduced density matrix for the state $\ket{\Psi}$ is
\begin{equation}
\begin{split}
\rho_{A}&=\Tr_{B}\dyad{\Psi}, \\ &= \Tr_{B}[p_{1}\dyad{\psi_{1}}+p_{2}\dyad{\psi_{2}}+\sqrt{p_{1}p_{2}}(\op{\psi_{1}}{\psi_{2}}+\op{\psi_{2}}{\psi_{1}})], \\ &= \begin{pmatrix} p_{1}a^{2}+p_{2}c^{2} & \sqrt{p_{1}p_{2}}(ad+bc) \\ \sqrt{p_{1}p_{2}}(ad+bc) & p_{1}b^{2}+p_{2}d^{2} \end{pmatrix}.
\end{split}
\end{equation}
The condition that $\det \rho_{A} = 0$ gives
\begin{equation}
p_{1}=\frac{cd}{ab+cd}, \, \, p_{2}=\frac{ab}{ab+cd}.
\end{equation}
Since $\ket{\Psi}$ is a product state, the left side of the majorization relation becomes
\begin{equation}
\lambda(\phi) \otimes \lambda(\Psi) = \begin{pmatrix} \lambda(\phi) \\ 0
\end{pmatrix}
\end{equation}
in block format. The relation, then becomes
\begin{equation}
\begin{pmatrix} \alpha^{2} \\ \beta^{2} \end{pmatrix} \prec \frac{cd}{ab+cd}\begin{pmatrix} a^{2} \\ b^{2} \end{pmatrix} + \frac{cd}{ab+cd}\begin{pmatrix} c^{2} \\ d^{2} \end{pmatrix},
\end{equation}
giving the following upper bound on $\alpha$
\begin{equation}
\alpha \leq \frac{a^{2}cd+c^{2}ab}{ab+cd}.
\end{equation}
The bound states that if the states $\ket{\psi_{1}},\ket{\psi_{2}}$ are maximally entangled, then $\ket{\phi}$ is also maximally entangled and if $\ket{\psi_{1}}, \ket{\psi_{2}}$ are product than $\ket{\phi}$ is also a product state. Converting this bound into a bound for entanglement entropy is not trivial and numerical techniques will in general be needed but in terms of concurrence, the entanglement of the state $\ket{\phi}$ has a simple, elegant form
\begin{equation}
\mathcal{C}(\phi) = \sqrt{\mathcal{C}(\psi_{1})\mathcal{C}(\psi_{2})},
\end{equation}
it is simply the geometric mean of the entanglements of the states.

The same kind of majorization relation can be used to find a lower bound on the Schmidt coefficients of a preshared state in the case of a 4 state discrimination scheme where the states $\ket{\psi_{i}}$ to be discriminated form a complete orthonormal basis
\begin{equation}
\begin{split}
\ket{\psi_{1}}&=a\ket{00}+b\ket{11},\\
\ket{\psi_{2}}&=b\ket{00}-a\ket{11},\\
\ket{\psi_{3}}&=c\ket{01}+d\ket{10},\\
\ket{\psi_{4}}&=d\ket{00}-c\ket{11},
\end{split}
\end{equation}
where $a,b,c,d \in \mathds{R}$ and $a\geq b$, $c \geq d$. As before, the parties also share the state $\ket{\phi} = \alpha \ket{00} + \beta \ket{11}$ and $\alpha, \beta \in \mathds{R}$, $\alpha \geq \beta$. The entanglement transformation procedure is $\ket{\phi} \otimes \sum_{i}\sqrt{p_{i}}\ket{\psi_{i}} \rightarrow \ket{\psi_{i}}$ with probability $p_{i}$ and $i$ identified, using the idea explained above. The majorization relation is constructed as
\begin{equation}
\lambda(\phi) \otimes \lambda(\sum_{i}\sqrt{p_{i}}\ket{\psi_{i}})\prec \sum_{i}p_{i}\lambda(\psi_{i}).
\end{equation}
However, the method of finding the values of $p_{i}$ that make the determinant of the reduced density operator won't work here. Instead the state $\ket{\Psi}=\sum_{i}\sqrt{p_{i}}\ket{\psi_{i}}$ can be made product explicitly by construction. 

Let $\ket{\Psi} = \ket{A} \otimes \ket{B}$ where
\begin{equation}
\begin{split}
\ket{A}&=x_{A}\ket{0}+y_{A}\ket{1},\\
\ket{B}&=x_{B}\ket{0}+y_{B}\ket{1}.
\end{split}
\end{equation}
The coefficients $\sqrt{p_{i}}$ are given by 
\begin{equation}\sqrt{p_{i}} = \braket{\psi_{i}}{A\otimes B}\end{equation} where $\ket{A\otimes B}$ is a shorthand for $\ket{A} \otimes \ket{B}$. Using eq. (4.15) the coefficients for the set of states in eq. (4.12) are found as
\begin{equation}
\begin{split}
\sqrt{p_{1}} &= x_{A}x_{B}a+y_{A}y_{B}b, \\ \sqrt{p_{2}} &= x_{A}x_{B}b-y_{A}y_{B}a, \\ \sqrt{p_{3}} &= x_{A}y_{B}c+y_{A}x_{B}d, \\ \sqrt{p_{4}} &= x_{A}y_{B}d-y_{A}x_{B}c,
\end{split}
\end{equation}
which make the state $\ket{\Psi}=\sum_{i}\sqrt{p_{i}}\ket{\psi_{i}}$ a product state. The majorization relation in eq. (4.13) becomes 
\begin{equation}
\begin{pmatrix} \lambda(\phi) \\ 0 \end{pmatrix} \prec \sum_{i}p_{i}\lambda(\psi_{i})
\end{equation}
since $\ket{\Psi}$ is a product state. Therefore 
\begin{equation}
\lambda(\sum_{i}\sqrt{p_{i}}\ket{\psi_{i}}) = \begin{pmatrix} 1 \\ 0 \end{pmatrix}.
\end{equation}
Since it was assumed that $\alpha \geq \beta$, only the first term of the column matrix in the left hand side of eq. (4.17), the inequality transforms into
\begin{equation}
\alpha^{2} \leq a^{2}(x_{A}^{2}x_{B}^{2}+(1-x_{A}^{2})^{2}(1-x_{B}^{2})^{2}+c^{2}(x_{A}^{2}(1-x_{B}^{2})^{2}+(1-x_{A}^{2})^{2}x_{B}^{2})
\end{equation}
after the insertion of the normalization conditions for the states $\ket{A}$ and $\ket{B}$.

In order to find the lowest possible upper bound on $\alpha^{2}$, the right hand side of eq. (4.19) should be minimized. The minimization procedure is easy since the equation is linear in $x_{A}^{2}$ and $x_{B}^{2}$ and a linear function has its extrema at the endpoints. This fact gives the bound on $\alpha^{2}$ as 
\begin{equation}
\alpha^{2} \leq \min(a^{2},c^{2}).
\end{equation}
In other words, the state $\ket{\phi}$ should at least have the same amount of entanglement as the most entangled state in the set $\{\ket{\psi_{i}}\}$. Note that this bound is a lower bound for this task. The entanglement cost of such an entanglement preserving procedure can be higher than the bound found here which will be shown in the following part.

To find the lowest possible upper bound and the highest possible lower bound on the Schmidt coefficients of the preshared state $\ket{\phi}$ by using the method of making the state $\ket{\Psi}$ a product state, the construction of the product state $\ket{\Psi}$ is reconsidered. It is constructed as follows
\begin{equation}
\ket{\Psi}=\sum_{\mu}\ket{\psi_{\mu}}_{AB}\otimes \ket{\varphi_{\mu}}_{\bar{A}\bar{B}}.
\end{equation}
where $A,\bar{A},A^{\prime},\hdots$ etc. represent various Hilbert spaces on Alice's side and the same goes for Bob. The procedure, as stated above can be considered as an entanglement transformation if Alice and Bob are able to go through with the discrimination. The entanglement transformation is as follows
\begin{equation}
\ket{\phi}_{AB} \otimes (\sum_{\mu}\ket{\psi_{\mu}}_{AB}\otimes \ket{\varphi_{\mu}}_{\bar{A}\bar{B}}) \rightarrow \frac{\ket{\psi_{\mu}}_{AB}\otimes \ket{\varphi_{\mu}}_{\bar{A}\bar{B}}}{\norm{\varphi_{\mu}}}
\end{equation}
with the index $\mu$ identified with probability $p_{\mu}=\norm{\varphi_{\mu}}^{2}$, meaning that the discrimination is successful. The majorization relation for this process can be expressed as follows
\begin{equation}
\lambda(\phi)\otimes \lambda(\sum_{\mu}\ket{\psi_{\mu}}_{AB}\otimes \ket{\varphi_{\mu}}_{\bar{A}\bar{B}}) \prec \sum_{\mu}\norm{\varphi_{\mu}}^{2}\lambda(\frac{\psi_{\mu} \otimes \varphi_{\mu}}{\norm{\varphi_{\mu}}}).
\end{equation}
The same argument used in the preceding parts of this chapter applies here, to find the lowest possible upper bound, the state $\ket{\Psi} =  \sum_{\mu}\ket{\psi_{\mu}}_{AB}\otimes \ket{\varphi_{\mu}}_{\bar{A}\bar{B}}$ must be made a product state, in this case the state $\ket{\Psi}$ must be unentangled in the $A\bar{A}:B\bar{B}$ cut to ensure that the state $\ket{\Psi}$ is unentangled between Alice and Bob.

A new type of product between kets and bras will be defined here
\begin{equation}
\ket{\varphi_{\mu}} = \braket{\psi_{\mu}}{\Psi}\rangle .
\end{equation}
If the state $\ket{\Psi}$ is unentangled in the $A\bar{A}:B\bar{B}$ cut, it can be expressed as $\ket{\Psi}=\ket{u}_{A\bar{A}}\otimes \ket{v}_{B\bar{B}}$. In order for this to be accomplished, the states $\ket{u}_{A\bar{A}}$ and $\ket{v}_{B\bar{B}}$ must be maximally entangled, this fact stems from the monogamy of entanglement. Since all maximally entangled states are unitarily equivalent to each other under LOCC, the states $\ket{u}, \, \ket{v}$ can be chosen as one of the bell states, choose them as
\begin{equation}
\begin{split}
\ket{u} &= \frac{1}{\sqrt{2}}(\ket{00}+\ket{11}) = \frac{1}{\sqrt{2}}(\sum_{i,j=0}^{1}\delta_{ij}\ket{ij}_{A\bar{A}}), \\ \ket{v} &= \frac{1}{\sqrt{2}}(\ket{00}+\ket{11}) = \frac{1}{\sqrt{2}}(\sum_{i,j=0}^{1}\delta_{ij}\ket{ij}_{B\bar{B}}).
\end{split}
\end{equation}
Using eq. (4.25), the state $\ket{\Psi}$ can be re-expressed as
\begin{equation}
\ket{\Psi}=\frac{1}{2}\sum_{i,j,k,l=0}^{1}\delta_{ij}\delta_{kl}\ket{ik}_{AB}\otimes \ket{jl}_{\bar{A}\bar{B}}.
\end{equation}
The states $\ket{\psi_{\mu}}$ can be expressed generally as
\begin{equation}
\ket{\psi_{\mu}}=\sum_{i,k}(\psi_{\mu})_{ik}\ket{ik}_{AB}.
\end{equation}
Then, the states $\ket{\varphi_{\mu}}$ can be found as
\begin{equation}
\begin{split}
\ket{\varphi_{\mu}}=\braket{\psi_{\mu}}{\Psi}\rangle &= \frac{1}{2}\sum_{i,j,k,l}(\psi_{\mu})^{\star}_{ik}\delta_{ij}\delta_{kl}\ket{jl}_{\bar{A}\bar{B}}, \\
&=\frac{1}{2}\sum_{i,k}(\psi_{\mu})^{\star}_{ik}\ket{ik}_{\bar{A}\bar{B}}, \\ 
\ket{\varphi_{\mu}} &= \frac{1}{2}\ket{\psi^{\star}_{\mu}}.
\end{split}
\end{equation}
Finally, the state $\ket{\Psi}$ becomes
\begin{equation}
\ket{\Psi}=\frac{1}{2}\sum_{\mu}\ket{\psi_{\mu}}_{AB} \otimes \ket{\psi_{\mu}^{\star}}_{\bar{A}\bar{B}}.
\end{equation}
Inserting the expression for $\ket{\varphi_{\mu}}$, the majorization relation in eq. (4.23) becomes
\begin{equation}
\lambda(\phi) \prec \frac{1}{4}\sum_{\mu}\lambda(\psi_{\mu})\otimes \lambda(\psi_{\mu})
\end{equation}
after using the facts that;

(i) $\norm{\varphi}=\frac{1}{2}$,

(ii) $\ket{\Psi}$ is a product state hence $\lambda(\Psi) = \begin{pmatrix} 1 \\ 0 \\ \vdots \\ 0 \end{pmatrix}$,

(iii) $\lambda(\psi_{\mu}) = \lambda(\psi_{\mu}^{\star})$.

Assuming that $a \geq b \geq 0$ and $c \geq d \geq 0$,
\begin{equation}
\begin{split}
\lambda(\psi_{1})=\lambda(\psi_{2}) &= \begin{pmatrix} a^{2} \\ b^{2} \end{pmatrix}, \\ \lambda(\psi_{3})=\lambda(\psi_{4}) &= \begin{pmatrix} c^{2} \\ d^{2} \end{pmatrix},
\end{split}
\end{equation}
the majorization relation can be written as
\begin{equation}
\lambda(\phi) \prec \frac{1}{2}\begin{pmatrix} a^{4} + c^{4} \\ a^{2}b^{2}+c^{2}d^{2} \\ a^{2}b^{2}+c^{2}d^{2} \\ b^{4} + d^{4} \end{pmatrix}.
\end{equation}
It can be inferred from the above equation that the state $\ket{\phi}$ should have a Schmidt rank of at least 4 if $a,b,c,d$ are all different from 0, otherwise the relation is not satisfiable by any means. The limiting cases that all the states are maximally entangled and all the states are product give the following results: $\ket{\phi}$ possesses 2 ebits of entanglement for the first case and $\ket{\phi}$ is a product state for the second case. The 2 ebit result is interesting but the explanation is quite simple. Suppose Alice and Bob, instead of realizing the discrimination by an entanglement transformation scheme opt to use quantum teleportation. Alice can teleport her part of the states to Bob using 1 ebit of entanglement and than Bob can implement a projective measurement to realize the discrimination of states. But, after the discrimination, the states are no longer shared between Alice and Bob and in order to preserve the states exactly as they were before, Bob needs to teleport the state back to Alice using 1 ebit. In total, it costs 2 ebits to discriminate the state while preserving it. 

\begin{figure}[h]
\begin{center}

\includegraphics[width=10cm]{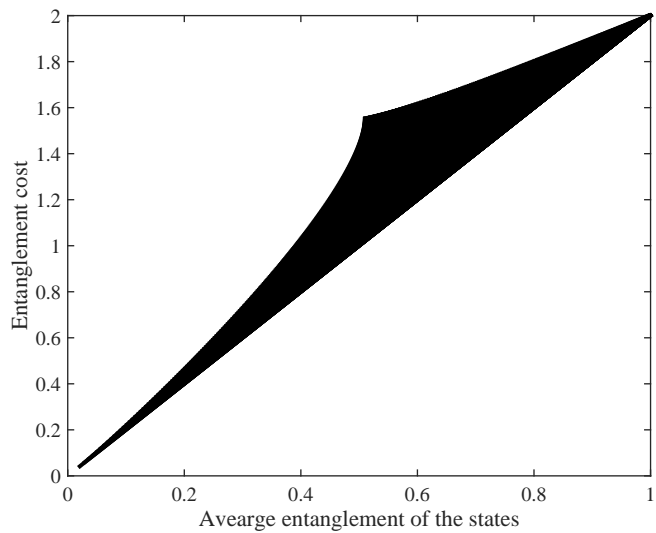}
\caption{Plot of the entanglement cost versus the average entanglement of the states. The diagonal line is when all the states have equal entanglement and the cusp is when 2 of the states are maximally entangled and the others are product states. The entanglement cost is not a simple function of the average entanglement as the average entanglement is not a one-to-one function of the parameters $a$ and $c$.\label{fig:avent4}}
\end{center}

\end{figure}

The minimum amount of entanglement that the state $\ket{\phi}$ possesses for successful discrimination can be calculated using the Schmidt coefficients as follows
\begin{equation}
\mathcal{E}(\phi) = -\sum_{i=1}^{4}\lambda_{i}\log_{2}\lambda_{i}
\end{equation}
but the expression is not one-to-one since different values of $a,c$ will always yield different values for the entanglement amount of the state $\ket{\phi}$ but the same amount of average entanglement of the states to be discriminated. Instead, a numerical calculation can be performed to plot the entanglement cost with respect to the average entanglement of the states assuming that the states have equal prior probabilities and it is shown in figure \ref{fig:avent4}.

In this work, it is claimed that the bound on the Schmidt coefficients of $\ket{\phi}$ in eq. (4.32) is the best possible bound to be found using majorization to characterize the discrimination process.

This last result concludes the discussion of discrimination with remaining entanglement in this thesis. The work done here is in no ways a general calculation as the states to be discriminated are all in very specific forms and the protocols to achieve discrimination are not constructed explicitly. Some problems arise during the generalization of the calculations performed in this chapter. For Hilbert spaces having a dimensionality greater than 2, the expressions for Bell-like maximally entangled states and complete bases tend to get complicated very quickly. The results found here are presented to show that the idea is indeed possible and to introduce the readers to the usefulness of majorization in state discrimination problems. Majorization is a very powerful tool and it completely characterizes any pure state entanglement manipulation under LOCC.

\section{Discussion and Conclusion}
There have been many achievements in the field of quantum state discrimination, especially for problems involving monopartite states. To this date, nearly all cases of monopartite state discrimination have been researched and solutions for specific cases have been found and optimized although to the knowledge of the author, a general, optimized solution hasn't been found yet. On the multipartite state discrimination front, necessary conditions that the states to be discriminated have been found\cite{chefcond} and the class of problems not solvable by LOCC but solvable by separable measurements have been characterized recently\cite{cohenclass}. The idea of preserving entanglement after discrimination is somewhat new and some progress have been made. Cohen, in his papers\cite{cohenent,cohenresource} introduced the idea but haven't calculated the amount of entanglement needed to achieve discrimination. On a very related subject, authors such as Bandyopadhyay \cite{2qubitcost,entcost} has calculated the entanglement cost of non-local measurements, which can also be used to achieve perfect discrimination of entangled states with remaining entanglement after a teleportation or, if a number of states can be discriminated with remaining entanglement, a projective measurement projecting onto those states can be constructed. In this thesis, characterization of such procedures using majorization was performed and the exact bounds on the Schmidt coefficients of the preshared entangled state were found. This is the original part in this thesis. 

The problem of discriminating 2 or 3 monopartite states was solved exactly in the literature during the early 2000s. The main ideas and the solutions presented in the opening sections of the third chapter of this thesis are for 2 state problems but they can easily be generalized to 3 states. Using the QSS procedure, which generalizes UQSD, solutions for 3 state problems can be found but as the number of states to be discriminated increase, finding the optimal solution gets increasingly harder. As of now, there is no general, optimal solution for the state discrimination problem neither for UQSD nor for minimum error discrimination. 

Also, in chapter three, majorization was introduced and it was shown that majorization can characterize state discrimination problems after the problems are reformulated as entanglement transformation problems. The most widely used and central majorization relation is
\begin{equation}
\lambda(\psi) \prec \sum_{i}p_{i}\lambda(\phi_{i}),
\end{equation} 
which governs a probabilistic entanglement transformation $\ket{\psi} \rightarrow \ket{\phi_{i}}$, with probability $p_{i}$. As a quick example, majorization was used to show indistinguishability of 4, orthogonal entangled qubit states with ease. 

Majorization is a great tool to analyze processes which use entanglement as a resource. To that end, in chapter three, it was shown that, using up some entanglement, 2 parties can distinguish between 4 orthogonal, entangled qubit states. In all the cases examined in this thesis, protocols to achieve the process were not explicitly constructed due to difficulty of generalizing LOCC measurements. 

The main aim of this thesis is to show that discrimination while preserving entanglement is possible. In chapter four, several cases of discrimination with remaining entanglement are inspected and it was shown that it is only possible if the parties agree upon using up some entanglement. Using majorization, the bounds on the Schmidt coefficients for the pre-shared state is found. Transforming these bounds into an analytical bounds for the entanglement of the preshared state in terms of the Schmidt coefficients of the states to be discriminated is not always possible but numerical methods were used to see the behavior. 

As stated above, Bandyopadhyay and his collaborators have performed the entanglement cost of nonlocal measurements using a different method. The result that was found in this thesis differs from their result but the explanation for the difference is straightforward. In their work, the procedure does not care about the preservation of the entanglement of the states thus the cost of a projective measurement projecting onto entangled states can be realized by a single teleportation at the worst case. As it was shown in the previous chapter, the cost of preserving entanglement after discrimination is always higher than the cost of a nonlocal measurement because of the need of another teleportation to preserve the original form of the states. 

In this thesis, the close relationship of transformation of pure entangled states and quantum state discrimination was also shown. The relation is easy to see for the monopartite case, the QSS generalization of UQSD is just a probabilistic state transformation procedure. For multipartite states, the general idea is the same and it is because of this connection that the theory of majorization and Nielsen's theorem can be applied to multipartite state discrimination problems to completely characterize the discrimination procedure. 

In conclusion, in this thesis, the state discrimination problem in quantum mechanics was presented and different cases of the problem were examined. The theory of majorization and its connections with the discrimination problem of multipartite states were also shown. Discrimination with entanglement preservation was also investigated and the possibility of such a procedure was shown and for some special cases, the procedure was completely characterized using majorization relations. 
\section*{Acknowledgments}
I am very grateful to my tutor and advisor, Prof. Dr. Sadi Turgut for his supervision, patience, encouragement and for the fruitful discussions throughout this study. He has introduced me to the beautiful subject of quantum information theory and his insight has proven invaluable on lots of occasions.

%
\bibliography{arxiv_ver1.bib}
\bibliographystyle{unsrt}
\bibliographystyle{unsrt}
\pagebreak
\appendix
\section{Probability Theory Fundamentals}
Notions of probability theory are used extensively in quantum mechanics and quantum information theory. Some of the relations in this appendix have been used widely in this thesis and they should be introduced or proven. The aim of this short appendix is to introduce some basic rules, definitions and relations of probability theory.

The fundamental object in the theory of probability is the random variable. A random variable is defined in the following way, a random variable $X$ takes the value $x$ with probability $p(X=x)$. $p(X=x)$ will be called just $p(x)$ as a shorthand notation when the meaning is clear. 

The joint probability notion is important and it is usually shown as $p(X=x,Y=y)$, or in shorthand $p(x,y)$ and it is the probability that $X=x$ and $Y=y$.

An important concept in probability theory is the conditional probability. It means the probability that $Y=y$ given $X=x$ and it is defined as
\begin{equation}
p(y|x)=\frac{p(x,y)}{p(x)}.
\end{equation}

A very widely used rule in probability theory is the Bayes' rule. It relates the conditional probabilities $p(x|y)$ and $p(y|x)$ with the formula
\begin{equation}
p(x|y)=\frac{p(y|x)p(x)}{p(y)}.
\end{equation}
The derivation is quite simple and uses the relation for joint probability
\begin{equation}
\begin{split}
p(x,y)=p(x)p(x|y)&=p(y)p(y|x), \\ p(x)p(x|y) &= p(y)p(y|x), \\ p(x|y)=\frac{p(y|x)p(x)}{p(y)}.
\end{split}
\end{equation}
Another widely used and important relation in the theory of probability is the law of total probability. It expresses the probability of $Y=y$ in terms of the probability of $X=x$ and $Y=y$ given $X=x$ with the following relation
\begin{equation}
p(y)=\sum_{x}p(y|x)p(x).
\end{equation}
To see this result, the joint probability expression must be used. For different values of $X$, the joint probability can be written as
\begin{equation}
\begin{split}
p(y,x_{1})&=p(y|x_{1})p(x_{1}), \\ p(y,x_{2})&=p(y|x_{2})p(x_{2}), \\ &\vdots
\end{split}
\end{equation}
Note that, if all these probabilities are summed for all possible values of $X$, it gives the joint probability of $Y=y$ and $X=x_{1},x_{2},\hdots,x_{N}$ which is just the probability $p(y)$.

The mean of a random variable, usually called the expectation value in quantum mechanics is the average of all the values a random variable $X$ can have weighted with respect to the probabilities $p(x)$ and it is equal to
\begin{equation}
\langle X \rangle = \sum_{x}p(x)x .
\end{equation}
This short review of probability theory is enough for the purposes of this work. The interested reader can refer to one of the many textbooks on this subject.
\end{document}